\documentclass{article}

\usepackage[table]{xcolor}

\usepackage[utf8]{inputenc}
\usepackage[T1]{fontenc}
\usepackage{url}
\usepackage{booktabs}
\usepackage{amsfonts}
\usepackage{nicefrac}
\usepackage{microtype}
\usepackage{lipsum}

\usepackage{amsmath}
\usepackage{amsthm}
\usepackage{amssymb}
\usepackage{mathtools}
\usepackage{bm}
\usepackage{bigints}

\usepackage{graphicx}
\usepackage{tikz-cd}

\usepackage{siunitx}
\usepackage{diagbox}
\usepackage{makecell}
\usepackage{multirow}
\usepackage{caption}
\usepackage{subcaption}

\usepackage[ruled]{algorithm2e}
\usepackage{setspace}

\usepackage{enumitem}

\usepackage{cite}

\usepackage{PRIMEarxiv}

\usepackage{hyperref}
\hypersetup{
    colorlinks=true,
    citecolor=black,
    linkcolor=black,
    filecolor=black,
    urlcolor=black,
}

\usepackage{pifont}

\newcommand{\R}{\mathbb{R}}

\renewcommand{\emptyset}{\varnothing}

\def\0{{\mathbb 0}}

\def\R{{\mathbb R}}

\def\curlyD{{\mathcal D}}
\def\curlyE{{\mathcal E}}

\def\curlyX{{\mathcal X}}
\def\curlyY{{\mathcal Y}}

\numberwithin{equation}{section}

\title{Micrometer: Micromechanics Transformer for Predicting Mechanical Responses of Heterogeneous Materials
}

\author{
  Sifan Wang\thanks{These authors contributed equally.} \\
Institution for Foundation of Data Science \\
  Yale University \\
  New Haven, CT 06520 \\
  \texttt { sifan.wang@yale.edu} \\
  \And
   Tong-Rui Liu\footnotemark[1] \\
  Department of Aeronautics \\ 
  Imperial college London\\
  London, SW7 2AZ\\
  \texttt {tongrui.liu18@imperial.ac.uk} \\
  \And
    Shyam Sankaran \\\
  Department of Mechanical Engineering  \\
  and Applied Mechanics\\
  University of Pennsylvania \\
  \texttt{shyamss@seas.upenn.edu} \\
  \And
  Paris Perdikaris \\
  Department of Mechanical Engineering \\
  and Applied Mechanics\\
  University of Pennsylvania\\
  Philadelphia, PA 19104 \\
  \texttt{pgp@seas.upenn.edu} 
}

\begin{document}
\maketitle

\begin{abstract}
Heterogeneous materials, crucial in various engineering applications, exhibit complex multiscale behavior, which challenges the effectiveness of traditional computational methods. In this work, we introduce the Micromechanics Transformer ({\em Micrometer}), an artificial intelligence (AI) framework for predicting the mechanical response of heterogeneous materials, bridging the gap between advanced data-driven methods and complex solid mechanics problems.  
Trained on a large-scale high-resolution dataset of 2D fiber-reinforced composites, Micrometer can achieve state-of-the-art performance in predicting microscale strain fields across a wide range of microstructures, material properties under any loading conditions and 
We demonstrate the accuracy and computational efficiency of Micrometer through applications in computational homogenization and multiscale modeling, where  Micrometer achieves 1\% error in predicting macroscale stress fields while reducing computational time by up to two orders of magnitude compared to conventional numerical solvers. We further showcase the adaptability of the proposed model through transfer learning experiments on new materials with limited data, highlighting its potential to tackle diverse scenarios in mechanical analysis of solid materials.
Our work represents a significant step towards AI-driven innovation in computational solid mechanics, addressing the limitations of traditional numerical methods and paving  the way for more efficient simulations of heterogeneous materials across various industrial applications.

\end{abstract}

\section{Introduction}

% big background of AI4science 

% background of AI for scientifc computing

% challenge and gap in applying AI for computaional mechanics
% 1. good dataset are in computaional fluid mechanics and most AI models foucs on time-dependent dynamics
% 2. lack of large scale high quality data in solid mechanics, 
% 3. Use simple small scale model to predict either macro mechanical responses or micro stress under resctriected conditions or range of materials, poor performance and low accuarcy

Heterogeneous materials, such as concretes, composites, and metals, exhibit intrinsically hierarchical structures with a wide range of behaviors and morphologies across various time and length scales \cite{Concretebook,Miltoncomposite}. These materials serve as crucial constituents in infrastructural components across diverse industries, including aerospace, marine, mechanical, and bioengineering \cite{JONES2012xiii,Ashbydesign}. Accurately predicting the mechanical responses of materials is the cornerstone for industrial-scale engineering design and material characterization \cite{Olsondesign}.

Traditional approaches, including analytical and experimental methods, often face limitations when dealing with the complex and multiscale nature of how materials respond to loading \cite{Fishnaturereview}. Over the past few decades, many computational methods have been formulated to address these challenges, offering robust and accurate numerical simulations. These include the finite element method (FEM) \cite{HughesbookFEM,WriggersFEM}, the spectral Fast Fourier Transform (FFT) method \cite{Moulinec1998FFT,Muller1996}, and the virtual element method \cite{VEMGuide,basicVEM}, among others. These computational approaches involve solving partial differential equations (PDEs) governed by principles in continuum solid mechanics, such as stress balance equations and Lippmann-Schwinger equations, without relying on empirical laws or laborious experimental work.

While traditional numerical methods have demonstrated their effectiveness in solving PDEs in computational solid mechanics, they face several drawbacks that hinder their real-world applications. First, the solution variables heavily rely on the resolution of spatial and temporal discretization, with physical field information corresponding only to fixed discretization instances. This limitation results in reduced flexibility, as field information at new query positions cannot be directly accessed. Second, the solution fields typically correspond to a single set of parameters, such as PDE coefficients, domain configuration, and boundary conditions. This constraint leads to inefficiencies when dealing with practical problems that require frequent PDE solving with different setups, such as material and structural design under uncertainty \cite{XiaoyuZhengNC2023,ZhengNC2023}, topology optimization \cite{hengyangli2019,ZhengCMAME2021}, and concurrent multiscale modeling \cite{Fishbookmultiscale,YvonnetCMAME2013UQ,FeyelChaboche2000FE2}.

To address these issues, recent advancements in  operator learning have opened up new avenues for solving PDEs by constructing approximations of maps between function spaces \cite{kovachki2024operator}.  Compared to traditional numerical methods, operator learning bears several advantages. On one hand, the prediction of neural operators is mesh/discretization invariant, which means that inference can be made on any location inside the computational domain with arbitrary discretization resolutions. On the other hand, as an infinite-dimensional PDE solution operator is learned, the inference for new sets of PDE parameters only requires fast evaluations of the neural operator without the effort of retraining.
Notable architectures include the Fourier Neural Operators (FNO) \cite{li2021fourier, guibas2021adaptive, wen2022u, fanaskov2023spectral,you2022learning} and Deep Operator Networks (DeepONet) \cite{lu2021learning,wang2022improved, wang2021learning, wang2023long, seidman2022nomad, zhu2023fourier, goswami2022physics}, which have shown impressive capabilities in learning complex, nonlinear operators with high efficiency. These AI-driven approaches not only complement traditional numerical methods, but often surpass them in efficiency and accuracy for certain classes of problems, introducing a promising path towards accelerating scientific simulation and discovery.

Despite these promising advancements in AI for scientific computing, significant challenges and gaps remain, particularly in the field of computational solid mechanics. In computational fluid dynamics, high-quality and comprehensive datasets are continuously curated (e.g., PDEBench \cite{takamoto2022pdebench}, PDEArena \cite{gupta2022towards}, CFDBench \cite{luo2023cfdbench}), fostering active research in adapting and developing machine learning models for these applications \cite{lin2021seamless, li2022fourier, li2024geometry, alkin2024universal, wen2022u}. This progress has already  led to the development of competitive specialized models such as DPOT \cite{hao2024dpot}, MPP \cite{mccabe2023multiple}, Poseidon \cite{herde2024poseidon}, and UPT \cite{alkin2024universal}.
In contrast, while the field of computational solid mechanics has begun to explore the potential of modern AI advancements, such as transformers  and diffusion models \cite{bastek2023inverse,buehler2022modeling,buehler2022end,lew2023single,buehler2023melm}, their developments are still in early stages,  
and are often constrained by small-scale datasets and limited to predicting the mechanical responses at either macroscale or microscale under restricted geometric configurations, loading conditions and material types/properties \cite{yang2021deep, ChenyangEFM,BradyCMAME2024,BradyCPB2022,BradyMOM2023,SouvikiSCI2022,SouvikJMPS2023}. Here we attribute the limited success that AI methods have so far enjoyed in solid mechanics to the lack of large-scale, high-quality datasets, partly due to the inherent complexity and computational cost of generating such data \cite{BessaDATABASE,Stefanreview2024}. These limitations significantly hinder the development of robust, generalizable AI solutions for complex solid mechanics problems, particularly those involving heterogeneous materials and multiscale phenomena.

To address the aforementioned challenges and capability gaps, we introduce {\em Micrometer} for predicting the mechanical response of heterogeneous materials. Our key and original contributions are summarized as follows:
\begin{itemize}[leftmargin=10mm]

    \item We formulate a novel operator learning problem of mapping the fourth-order elastic tensor to the strain concentration tensor governed by the parametric Lippmann-Schwinger equation, enabling efficient evaluation of mechanical responses across varying microstructures and material properties, under any external loading conditions.

    \item To overcome the data scarcity, we generate and release CSMBench/CMME; the first large-scale, high-fidelity, high-resolution dataset in micromechanics, which covers a vast range of volume fractions, microstructural morphologies and material properties.
    
    \item  We develop a transformer-based AI model to learn the target operator, outperforming several popular PDE surrogates and exhibiting strong scalability with respect to data size, model parameters, and computational resources.
        
    \item We demonstrate the accuracy and computational efficiency of the proposed model by applying it to fundamental problems in micromechanics, including  computational homogenization and concurrent multiscale modeling. 
    
    \item We showcase the adaptability of our model through transfer learning experiments on new microstructures with limited data, highlighting its potential to tackle diverse scenarios in computational solid mechanics.
\end{itemize}

Taken  all together, our work  represents a significant step towards AI-driven innovation in computational mechanics and provides a crucial resource for future research in this field.

% % summary of the  paper
The structure of this paper is as follows
Section \ref{math_theory} introduces the essential mathematical preliminaries, establishing the theoretical foundation for our work. In Section \ref{sec: problem_formulation}, we formulate the problem of interest, identifying the appropriate target solution operator and defining the corresponding input and output function spaces. We then details the pipeline for generating the CSMBench/CMME dataset, crucial for training and evaluating our model.
Section \ref{sec: method} elucidates the general framework of operator learning and introduces our proposed  model, {\em Micrometer}. In Section \ref{sec: results}, we validate  Micrometer's performance through comprehensive comparisons against other state-of-the-art AI baselines. Next, we showcase its robustness and computational efficiency in addressing fundamental problems in computational solid mechanics, including  homogenization and concurrent multiscale modeling. Furthermore, we illustrate the adaptability of our model through fine-tuning experiments on out-of-distribution scenarios with limited data.
Finally, in Section \ref{sec: discussion},  we conclude by summarizing our key findings, discussing the current limitations of our approach, and outlining promising directions for future research.

% \section{Problem Setup}
\label{setup}
\section{Mathematical preliminaries}
\label{math_theory}

% introduction on FE-FFT

% introduction to micro scale problem
This section provides an overview of micro-elasticity in computational micromechanics, which involves solving the Lippmann-Schwinger equation in a 2D representative volume element (RVE). Let $\Omega$ be a computational domain of a RVE in 2D, the quasi-static linear elasticity problem in the absence of body forces can be formulated as 
\begin{equation}
\begin{dcases*}
\frac{\partial\sigma_{i j}(\mathbf{x})}{\partial x_i}=0, \\
\sigma_{i j}(\mathbf{x})=C_{i j k l}(\mathbf{x}): \varepsilon_{k l}(\mathbf{x}), \\
\varepsilon_{i j}(\mathbf{x})=\frac{1}{2}\left(\frac{\partial s_i(\mathbf{x})}{\partial x_j}+\frac{\partial s_j(\mathbf{x})}{\partial x_i}\right).
\end{dcases*}
\label{setup:balanceequation}
\end{equation}
The above relations represent  equilibrium, constitutive, and infinitesimal strain-displacement equations, respectively. Herein, $\sigma_{i j}, \varepsilon_{i j}, C_{i j k l}$ and $s_{i}$ denote stress, strain, fourth-order elastic tensor and displacement, respectively. Since the RVE is intrinsically heterogeneous, local constitutive relations and physical fields are denoted as functions of position $\mathbf{x} \in \Omega$. The local elastic tensor $C_{ijkl}(\mathbf{x})$ can be expressed as a pair of Lam\'e constants as
\begin{equation}
C_{i j k l}(\mathbf{x})=\lambda(\mathbf{x}) \delta_{i j} \delta_{k l}+\mu(\mathbf{x})\left(\delta_{i k} \delta_{j l}+\delta_{i l} \delta_{j k}\right),
\label{setup:Lametransform}
\end{equation}
where $\lambda(\mathbf{x})$ and $\mu(\mathbf{x})$ can be computed through Young's modulus $E(\mathbf{x})$ and Poisson ratio $\nu(\mathbf{x})$ as
\begin{align}
    \lambda(\mathbf{x}) = \frac{E(\mathbf{x})\nu(\mathbf{x})}{(1+\nu(\mathbf{x}))(1-2\nu(\mathbf{x}))},  \quad \mu(\mathbf{x}) = \frac{E(\mathbf{x})}{2(1+\nu(\mathbf{x}))}.
    \label{setup:Lameconstant}
\end{align}
To close the above system, we impose periodic boundary conditions $(\mathrm{PBC}$) for the RVE. Moreover, the point-wise displacement field can be decomposed as
\begin{equation}
s_{i}(\mathbf{x})=\tilde{s}_{i}(\mathbf{x})+\bar{s}_{i}.
\end{equation}
Without considering strain gradient effects and taking the symmetric gradient of the above equation, the corresponding strain field can be split as
\begin{equation}
\varepsilon_{i j}(\mathbf{x})=\tilde{\varepsilon}_{i j}(\mathbf{x})+\bar{\varepsilon}_{i j},
\label{setup:polarstrain}
\end{equation}
where $\tilde{s}_{i}(\mathbf{x}), \tilde{\varepsilon}_{i j}(\mathbf{x}), \bar{s}_{i}$ and $\bar{\varepsilon}_{i j}$ denote the displacement fluctuation, strain field fluctuation, average displacement, and average displacement gradient, respectively. Due to the nature of periodic boundary conditions, $\tilde{s}_{i}(\mathbf{x})$ and $\tilde{\varepsilon}_{i j}(\mathbf{x})$ are periodic, while the traction vector $\boldsymbol{\sigma} \cdot \boldsymbol{n}$ behaves anti-periodically on the boundary between adjacent RVEs. Here $\boldsymbol{n}$ denotes the outward normal vector between RVE boundaries. 

Next, we introduce a linear elastic homogeneous reference medium $C_{i j k l}^{0}$ \cite{Hashin1963HS,Hill1965SC} and express $C_{i j k l}(\mathbf{x})$ as
\begin{equation}
C_{i j k l}(\mathbf{x})=C_{i j k l}^{0}+\left[C_{i j k l}(\mathbf{x})-C_{i j k l}^{0}\right],
\end{equation}
\begin{equation}
C_{i j k l}^{0}=\lambda^{0} \delta_{i j} \delta_{k l}+\mu^{0}\left(\delta_{i k} \delta_{j l}+\delta_{i l} \delta_{j k}\right),
\label{setup:LameconstantC0}
\end{equation}
where $\lambda^{0}$ and $\mu^{0}$ are the Lam\'e constants of the elastic homogeneous reference material. Note that $\lambda^{0}$ and $\mu^{0}$ can be computed by the Young's modulus $E^{0}$ and Poisson ratio $\nu^{0}$ through Eq.\eqref{setup:Lameconstant}.
Then, the stress $\sigma_{ij}$ can be divided into two parts
\begin{equation}
\sigma_{i j}(\mathbf{x})=\tau_{i j}(\mathbf{x})+C_{i j k l}^{0}: \varepsilon_{k l}(\mathbf{x}),
\label{setup:polarstress}
\end{equation}
\begin{equation}
\tau_{i j}(\mathbf{x})=\left(C_{i j k l}(\mathbf{x})-C_{i j k l}^{0}\right): \varepsilon_{k l}(\mathbf{x}),
\end{equation}
where $\tau_{i j}(\mathbf{x})$ is a polarization stress that quantifies the difference between the stress in the real material and the stress in the reference material under the same strain. Combining Eq.\eqref{setup:polarstrain} and substituting Eq.\eqref{setup:polarstress} into the first equation in Eq.\eqref{setup:balanceequation}, we obtain
\begin{equation}
C_{i j k l}^{0} \frac{\partial \tilde{\varepsilon}_{i j}(\mathbf{x})}{\partial  x_i}=-\frac{\partial \tau_{i j}(\mathbf{x})}{\partial  x_i}.
\end{equation}
Because the second term in the right-hand side of Eq.\eqref{setup:polarstress} is linear, we can obtain the Lippmann-Schwinger equation \cite{Korner1971book} for strain field fluctuation in a convolution form by virtue of the Green's function $G_{i j k l}^{(0)}\left(\mathbf{x}, \mathbf{x}^{\prime}\right)$,
\begin{equation}
\varepsilon_{i j}(\mathbf{x})+G_{i j k l}^{(0)}\left(\mathbf{x}, \mathbf{x}^{\prime}\right) * \tau_{k l}\left(\mathbf{x}^{\prime}\right)-\bar{\varepsilon}_{i j}=0.
\label{setup:convolutionLS}
\end{equation}
The above equation establishes the relationship between the strain field fluctuation $\tilde{\varepsilon}_{i j} \left(\mathbf{x}\right)$ and polarization stress $\tau_{i j}(\mathbf{x})$ in Euclidean space. It can also be written in an integral form as below
\begin{equation}
\varepsilon_{i j}(\mathbf{x})+\int_{\Omega} G_{i j k l}^{(0)}\left(\mathbf{x}, \mathbf{x}^{\prime}\right): \tau_{k l}\left(\mathbf{x}^{\prime}\right) \mathrm{d} \mathbf{x}^{\prime}-\bar{\varepsilon}_{i j}=0.
\label{setup:dotLS}
\end{equation}

\section{Problem formulation}
\label{sec: problem_formulation}

Our primary objective is to develop a deep learning model for accurately predicting mechanical responses of heterogeneous materials.  Here we specifically focus on fiber-reinforced composites (e.g., carbon fiber, glass fiber), which are widely used in applications ranging from high-performance sport cars and aircrafts, to industrial structures like cooling towers. These materials play a crucial role in both everyday products and advanced engineering systems due to their exceptional strength-to-weight ratio and customizable properties.Nevertheless, the proposed framework can be readily extended to other types of heterogeneous materials as well.

We are interested in learning the  solution operator of the parameterized Lippmann-Schwinger equation Eq.\eqref{setup:dotLS}  by creating a mapping from various parameters, including material properties and microstructural configurations, to the mechanical responses of the material  under any external loading conditions. These parameters are represented by a fourth order elasticity tensor $\mathbb{C}\in L_\text{per}^\infty(\Omega,\mathbb{R}^{3\times3}_\text{sym})$, while the mechanical responses are characterized by a 
strain concentration tensor $\mathbb{A}\in L_\text{per}^2(\Omega,\mathbb{R}^{3\times3})$,
To this end, the operator mapping  can be defined as:
\begin{equation}
    \Phi: \mathbb{C} \in L_\text{per}^\infty \left(\Omega,\mathbb{R}^{3\times3}_\text{sym}\right) \longrightarrow  \mathbb{A} \in L_\text{per}^2\left(\Omega,\mathbb{R}^{3\times3} \right) 
\end{equation}
Here $\Omega$ represents the computational domain of RVE, and $\square_\text{per}$ is used to describe periodic functions. The rationale for selecting these input and output feature representations will be given as follows. 

\paragraph{Output feature representation.}
The mechanical responses of composite materials are typically characterized by the local strain field $\boldsymbol{\varepsilon}(\mathbf{x})$, which effectively captures the strain distribution pattern inside a RVE. However, selecting the strain field as a model target is not suitable due to its dependence on macrostrain boundary conditions $\overline{\boldsymbol{\varepsilon}}$. Consequently, the support of the distribution of $\boldsymbol{\varepsilon}(\mathbf{x})$ is unbounded if the distribution of $\overline{\boldsymbol{\varepsilon}}$ is unbounded,  making it impractical to generate sufficient data to cover the entire strain field distribution.

To address this difficulty,  we choose the point-wise strain concentration tensor $\mathbb{A}(\mathbf{x})$ as the \textit{output} of our model. Based on micromechanics theory (see Hill \cite{Hill1963}, Michel and Suquet \cite{Michel2003NTFA}, Yvonnet \cite{Yvonnet2019book}, Zohdi and Wriggers \cite{Zohdi2005book}), $\mathbb{A}(\mathbf{x})$ establishes  a linear transformation from the macrostrain boundary condition $\overline{\boldsymbol{\varepsilon}}$ to the local microstrain $\boldsymbol{\varepsilon}(\mathbf{x})$ as  
\begin{equation}
    \boldsymbol{\varepsilon}(\mathbf{x}) 
 = \mathbb{A}(\mathbf{x}): \overline{\boldsymbol{\varepsilon}}= \begin{bmatrix}
A_{11}(\mathbf{x}) & A_{12}(\mathbf{x}) & A_{13}(\mathbf{x})\\
A_{21}(\mathbf{x}) & A_{22}(\mathbf{x}) & A_{23}(\mathbf{x})\\
A_{31}(\mathbf{x}) & A_{32}(\mathbf{x}) & A_{33}(\mathbf{x})
\end{bmatrix}: \begin{bmatrix}
    \overline{\varepsilon}_{11}\\
    \overline{\varepsilon}_{22}\\
    \overline{\varepsilon}_{12}\\
\end{bmatrix}.
\label{eq:Atensor}
\end{equation}
Here, the point-wise strain concentration tensor $\mathbb{A}(\mathbf{x})$ is a $3 \times 3$ matrix independent of loading conditions for a RVE exhibiting linear elastic behavior. It can be obtained by solving the Lippmann-Schwinger equation subject to three orthogonal unit macroscopic strains $\overline{\varepsilon}_{ij}$ in each loading case
\begin{equation}
    \overline{\boldsymbol{\varepsilon}} = \begin{Bmatrix}\begin{bmatrix}
    1\\
    0\\
    0\\
\end{bmatrix},
\begin{bmatrix}
    0\\
    1\\
    0\\
\end{bmatrix},
\begin{bmatrix}
    0\\
    0\\
    1\\
\end{bmatrix}\end{Bmatrix}.
    \label{setup:macroBC}
\end{equation}
Importantly,  once the point-wise strain concentration tensor $\mathbb{A}(\mathbf{x})$ is determined, the local microstrain $\boldsymbol{\varepsilon}(\mathbf{x})$ can be directly computed by using the aforementioned formula for any given macrostrain boundary conditions $\overline{\boldsymbol{\varepsilon}}$. 
\paragraph{Input Feature Representation.}

When the RVE exhibits linear elastic behavior, the strain concentration tensor $\mathbb{A}(\mathbf{x})$ induces a unique mapping from macrostrain $\overline{\boldsymbol{\varepsilon}}$ to microstrain $\boldsymbol{\varepsilon}(\mathbf{x})$, independent of the loading conditions. Therefore,
the selection of $\mathbb{A}(\mathbf{x})$ not only reduces the effort in generating the output datasets, but also eliminates the requirements for taking loading conditions as \textit{inputs} of our model.
Accordingly, we choose the fourth-order elastic tensor $\mathbb{C}(\mathbf{x})$ as the input, which is the coefficient of Eq.\eqref{setup:dotLS} and is determined by the material properties and microstructural configuration of the RVE.

As illustrated in Section \ref{math_theory}, the domain $\Omega$ is discretized with pixels, and each pixel $\mathbf{x}$ is given by a set of material properties, including Young's modulus $E(\mathbf{x})$ and Poisson ratio $\nu(\mathbf{x})$. As the RVE is considered as a fiber reinforced composite (FRP), the domain $\Omega$ can be decomposed into disjoint fiber and matrix domains, denoted by ${\Omega}_f$ and ${\Omega}_m$, respectively,
such that ${\Omega}_m \cap  {\Omega}_f=\emptyset$ and  $\overline{{\Omega}_m \cup  {\Omega}_f}=\Omega$.
We then introduce a characteristic function $\chi(\mathbf{x})$ to denote the underlying microstructural configuration of the RVE as
\begin{equation}
\chi(\mathbf{x})= \begin{cases}
            1, & \text { if } \mathbf{x} \in \Omega_{f}, \\
            0, & \text { if } \mathbf{x} \in \Omega_{m}.
\end{cases}
\end{equation}
The function $\chi(\mathbf{x})$ determines the microstructural configuration of the RVE and is generated by a random fiber generation algorithm (see Section \ref{sec:data_generation}). 
Then, the point-wise Young's modulus and Poisson ratio can be expressed as
\begin{equation}
\begin{cases}
    E(\mathbf{x}) = \chi(\mathbf{x})E_f+(1-\chi(\mathbf{x}))E_m, \\
    \nu(\mathbf{x}) = \chi(\mathbf{x})\nu_f+(1-\chi(\mathbf{x}))\nu_m.
\end{cases}  
\end{equation}
%where $E_f, E_m, \nu_f, \nu_m$ are the Young's moduli and Poisson's ratios for the fiber and matrix materials, respectively. For a given RVE, these values are constant.  Given $E(\mathbf{x})$ and $\nu(\mathbf{x})$, we can compute the fourth-order elastic tensor $\mathbb{C}(\mathbf{x})$ using the relations in Eq.(\ref{setup:Lametransform}) and Eq.(\ref{setup:Lameconstant}).
%
Here, $E_{\square}$, $\nu_{\square} \in \R$ are a set of material properties including Young's modulus and Poisson ratio for domain $\Omega_{\square}$, where either the fiber domain $\Omega_{f}$ or the matrix domain $\Omega_{m}$ can be considered. $E_{\square}$ and $\nu_{\square}$ are unique for a given RVE, and the selections of these values will be discussed in Section \ref{sec:data_generation}. In addition, given $E(\mathbf{x})$ and $\nu(\mathbf{x})$, we can compute the fourth order elastic tensor $\mathbb{C}(\mathbf{x})$ using the relations in Eq.\eqref{setup:Lametransform} and Eq.\eqref{setup:Lameconstant}.

\begin{table}
\renewcommand{\arraystretch}{1.4}
\centering
\caption{ Summary of the CSMBench/CMME dataset parameters used in the numerical simulation of fiber-reinforced composite materials. }
\begin{tabular}{lc}
\toprule
\textbf{Dataset Size} & \textbf{Value} \\
\midrule
\# Training samples & 40,000 \\
\# Test samples & 2,000 \\
\midrule
\multicolumn{2}{l}{\textbf{RVE Characteristics}} \\
\midrule
RVE size $(\mu \text{m})$ &  $50 \times 50$ \\
Discretization resolution &  $512 \times 512$ \\
Fiber radius $R_d$ $(\mu \text{m})$  &  3.5 \\
Fiber radius standard deviation $\textit{Std}$ &  1\% \\
Fiber volume fraction $\textit{Vof}$ &  $40 \text{-} 60 \%$ \\
\midrule
\multicolumn{2}{l}{\textbf{Material Properties}} \\
\midrule
Fiber Young's modulus $E_f$ (\text{GPa}) & $5 \text{-} 85$\\
Fiber Poisson's ratio $\nu_f$ &  $0.05 \text{-} 0.45$ \\
Matrix Young's modulus $E_m$ (\text{GPa})&  $2.5 \text{-} 5$\\
Matrix Poisson's ratio $\nu_m$ & $0.3 \text{-} 0.4$ \\
\bottomrule
\label{tab:dataset}
\end{tabular}
\end{table}

\subsection{Data Generation}
\label{sec:data_generation}
Our data generation process is part of a broader effort coined as CSMBench (Computational Solid Mechanics Benchmark), which we envision as a comprehensive collection of datasets at the intersection of machine learning and computational solid mechanics. CSMBench aims to accelerate research and foster innovation by providing diverse, high-quality datasets for training and evaluating machine learning models in this rapidly evolving field.

Within CSMBench, here we focus on developing CMME (Computational Micromechanics for Elasticity) dataset, specifically designed to address linear elasticity problems in micromechanics. The CMME data generation process comprises two main components:  input data generation and output data generation. Each of these components plays a crucial role in creating a robust and representative dataset for our machine learning models. The main metadata of the CMME training and test datasets are summarized in Table \ref{tab:dataset}.

% \paragraph{RVE generation.}
The CSMBench/CMME dataset consists of 40,000 training samples and 2,000 test samples, each representing an input-output pair. The input consists of a periodic RVE domain with associated material parameters $E_f$, $\nu_f$, $E_m$, and $\nu_m$, while the output represents corresponding solutions of the Lippmann-Schwinger equation subjected to macrostrain boundary conditions stated in Eq.\eqref{setup:macroBC}.
For the process of RVE generation, we employ the $\texttt{RAND{\_}uSTRU{\_}GEN}$ algorithm proposed by Melro {\em et al.} \cite{Melrogenerate2008},  because of its ability to generate RVEs with high fiber volume fractions up to 65\%. We implement this algorithm using the open-source project  $\texttt{rvesimulator}$ \cite{yi2023rvesimulator}. 
To ensure a diverse and representative dataset, we stratify both the training and test samples into 20 groups, with fiber volume fractions uniformly distributed between 40\% and 60\%. The radius of fiber is set as 3.5$\mu \text{m}$ with 1\% standard deviation.
Each RVE domain size is set to be 50$\mu \text{m}$ $\times$ 50$\mu \text{m}$, discretized at a high resolution of 512 $\times$ 512 pixels.
For all training and test samples, the material properties $E_f$, $\nu_f$, $E_m$, and $\nu_m$ are generated via Latin hypercube sampling method subject the to upper and lower bounds provided in Table \ref{tab:dataset}.  
To compute the corresponding outputs, we solve Eq. \eqref{setup:convolutionLS} using a Fast Fourier Transform (FFT) based homogenization method \cite{Moulinec1998FFT}.  The details of the algorithm and the full pipeline of data generation are presented in  Appendix  \ref{appendix: fft}.

\section{Methods}
\label{sec: method}

\subsection{Operator learning}

Here we present an overview of  operator learning  \cite{chen2023deep,kovachki2024operator} that forms the basis for formulating our proposed  model.  Consider a general nonlinear operator $\Phi: \mathcal{X} \to \mathcal{Y}$ mapping between separable Banach spaces $\mathcal{X}$ and $\mathcal{Y}$. Our objective is to learn the operator $\Phi$ from a set of samples:
\begin{align}
    \label{eq: data}
    \left\{(u_n, v_n):  u_n \sim \mu, v_n =  \Phi\left(u_n\right)\right\}_{n=1}^N,
\end{align}
where $u_n, v_n$ are vector-valued functions in $\mathcal{X}$ and $\mathcal{Y}$, respectively, and the probability measure $\mu$ is supported on $\mathcal{X}$.

In practical implementations, since neural networks operate on finite-dimensional spaces, the foundation of many neural operators lies in the extraction of latent finite-dimensional structures, as illustrated below.
\[
\begin{tikzcd}[row sep=large, column sep=large]
\mathcal{X} \arrow[r, "\mathcal{E}_\curlyX"] \arrow[d, "\Phi"'] & \mathbb{R}^{d_\curlyX} \arrow[r, "\curlyD_\curlyX"] \arrow[d, "\Psi"] & \mathcal{X} \arrow[d, "\Phi"] \\
\mathcal{Y} \arrow[r, "\curlyE_\curlyY"] & \mathbb{R}^{d_\curlyY} \arrow[r, "\curlyD_\curlyY"]  & \mathcal{Y}
\end{tikzcd}
\]
Here, we introduce two encoder/decoder pairs on $\mathcal{X}$ and $\mathcal{Y}$. We denote $\curlyE_{\mathcal{X}}$ as the encoder mapping from the Banach space $\mathcal{X}$ to a Euclidean space $\mathbb{R}^{d_{\mathcal{X}}}$, where $d_{\mathcal{X}}$ denotes the encoding dimension. The corresponding decoder for $\mathcal{X}$ is defined as $\mathcal{D}_{\mathcal{X}}: \mathbb{R}^{d_{\mathcal{X}}} \to \mathcal{X}$.
Similarly, for the output space $\mathcal{Y}$, we define $\mathcal{E}_{\mathcal{Y}}: \mathcal{Y} \to \mathbb{R}^{d_{\mathcal{Y}}}$ as the encoder mapping from $\mathcal{Y}$ to a Euclidean space $\mathbb{R}^{d_{\mathcal{Y}}}$, with $d_{\mathcal{Y}}$ representing the encoding dimension for $\mathcal{Y}$. The corresponding decoder is defined as $\mathcal{D}_{\mathcal{Y}}: \mathbb{R}^{d_{\mathcal{Y}}} \to \mathcal{Y}$.

These encoder-decoder pairs approximately satisfy
\begin{align}
    \curlyE_{\mathcal{X}} \circ \curlyD_{\mathcal{X}} \approx I_{\mathcal{X}}, \quad \curlyE_{\mathcal{Y}} \circ \curlyD_{\mathcal{Y}} \approx I_{\mathcal{Y}},
\end{align}
where $I_\mathcal{X}$ and $I_\mathcal{Y}$ are the identity maps on $\mathcal{X}$ and $\mathcal{Y}$, respectively. 

We aim to determine an approximation to $\Phi: \mathcal{X} \to \mathcal{Y}$ by a family of parameterized functions 
\begin{align}
    \Psi: \R^{d_\mathcal{X}} \times \Theta \mapsto \R^{d_\mathcal{Y}},
\end{align}
where $\Theta \subseteq \mathbb{R}^p$ denotes the parameter space from which we seek the optimal choice of parameter, denoted $\theta^\star$. The approximation of $\Phi$ is achieved through the composition 
\begin{align}
    \curlyD_{\mathcal{Y}} \circ \Psi \circ \curlyE_{\mathcal{X}} \approx \Phi.
\end{align}
Here, the map $\curlyE_{\mathcal{X}}$ extracts a finite dimensional latent space from the input Banach space while the map $\curlyD_{\mathcal{Y}}$ returns from a second finite dimensional latent space to the output Banach space. Given the data in Eq.\eqref{eq: data}, these encoder-decoder pairs can be learned, reducing the operator approximation to a finite dimensional optimization problem 
\begin{align}
   \theta^* =   \underset{\theta \in \Theta}{\operatorname{argmin}} \frac{1}{N} \sum_{i=1}^N\left\|\Psi \circ \curlyE_{\mathcal{X}}\left(u_i\right)-\curlyE_{\mathcal{Y}}\left(v_i\right)\right\|_2^2.
\end{align}
This  formulation provides a general framework for learning complex nonlinear operators between Banach spaces, with applications spanning various domains in scientific computing and machine learning.

\subsection{The Micromechanics Transformer (Micrometer) }

\begin{figure}
    \centering
    \includegraphics[width=1.0\linewidth]{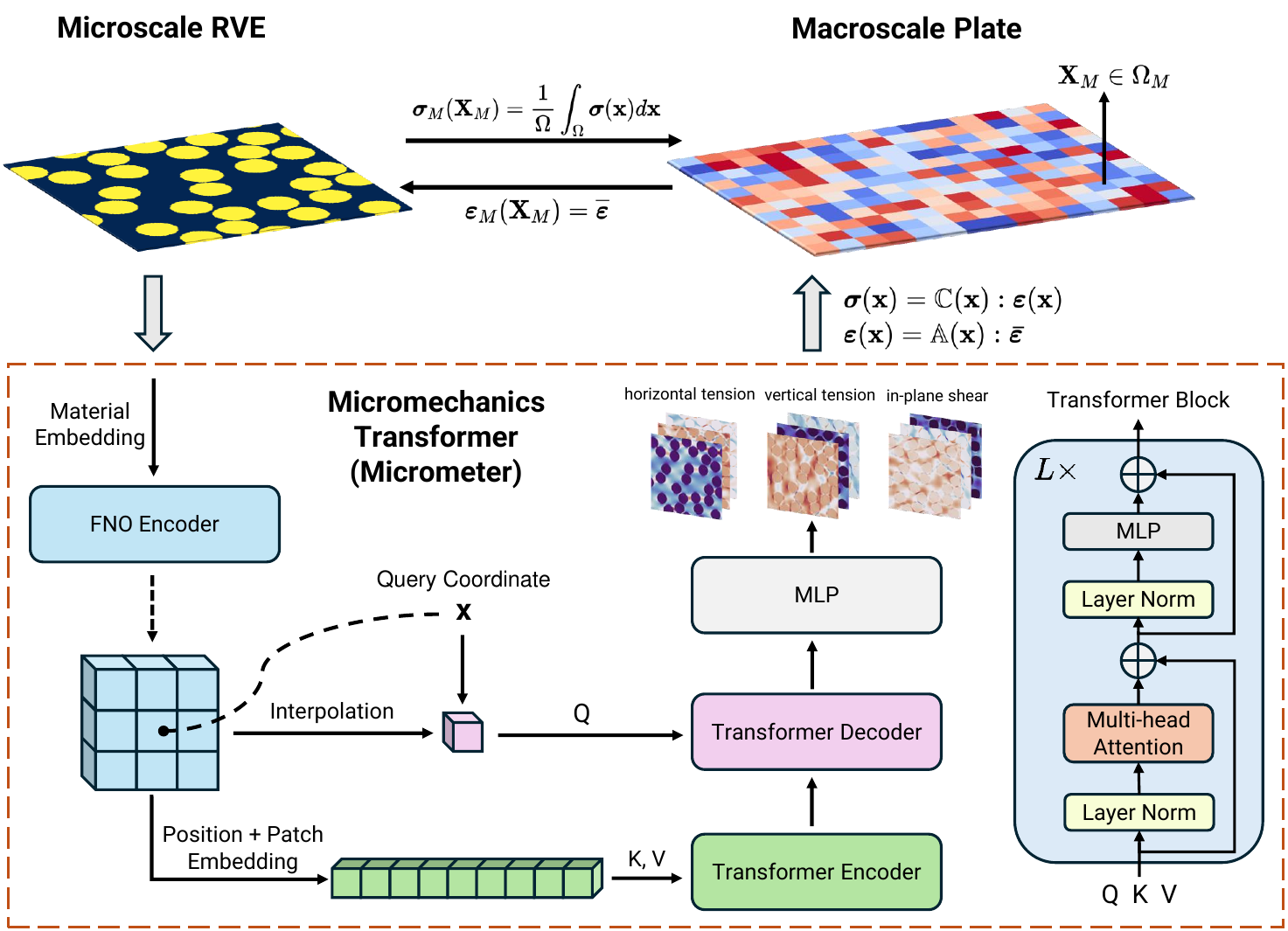}
    \caption{{\em Pipeline of Micrometer:} Micrometer is a transformer-based deep learning model used to predict the mechanical responses of fiber-reinforced composite materials. It takes representative volume elements (RVEs) and material properties at the microscale (e.g., Poisson's ratio, Young's modulus) as inputs, and outputs the corresponding point-wise strain concentration tensor governed by Lippmann-Schwinger equation.  
    Micrometer employs an encoder-decoder architecture. The input is first embedded using a resolution-invariant Fourier neural operator encoder. The resulting latent outputs are then patchified into a sequence of tokens, added with positional embeddings, and processed through a standard transformer backbone.
    A key feature of Micrometer is its ability to continuously evaluate the outputs at any query  coordinate. This is achieved by interpolating the FNO encoder outputs using Nadaraya-Watson kernel interpolation to obtain latent query-specific features. These features are then decoded by a standard transformer decoder with cross-attention between the query features and the outputs of the transformer encoder. Finally, a multilayer perceptron (MLP) is used to decode the output into the physical space. The pre-trained Micrometer model can be easily integrated with computational micromechanics frameworks, facilitating fast and accurate homogenization and multiscale modeling of composite materials. Furthermore, Micrometer can be further fine-tuned for a variety of downstream tasks with limited data.} 
    \label{fig:micrometer}
\end{figure}

Now we describe the network architecture of our proposed model, the  Micromechanics Transformer (Micrometer). As illustrated in Figure \ref{fig:micrometer}, this architecture consists of an encoder-decoder structure. The details of both the encoder and decoder components are described below.

\paragraph{Encoder.} As discussed in Section \ref{sec: problem_formulation}, we are interested in learning the solution operator of Lippmann-Schwinger equation where its input fourth order elasticity tensor $\mathbb{C}$ is parameterized by a microstructural configuration $\chi$ and material properties,  specifically the Young's modulus $E_f$, $E_m$ and Poisson's ratios $\nu_f$, $\nu_m$ for the fiber and matrix phases, respectively. Consequently, these parameters collectively serve as the  inputs to the encoder component of our model.

Let $\mathbf{a} \in \mathbb{R}^{H \times W \times 1}$ represent a discretization of a microstructure of interest, where 0 denotes the matrix and 1 denotes fiber. We remark that here $H, W$ denotes the resolution of the pixel grid, which is consistent with Eq. \eqref{setup:spa_discretized}. 
We begin by embedding the material properties, namely  the Young's modulus and Poisson's ratio of the matrix and fiber, into our inputs. These scalar properties are repeated for each pixel and concatenated with the original input, resulting in a tensor of shape $\mathbb{R}^{H \times W \times 5}$. Unlike conventional Vision Transformers (ViT), we first employ resolution-invariant Fourier Neural Operator (FNO) \cite{li2022fourier} layers (see Appendix \ref{app: fno}) to encode the inputs into a latent space, with its output denoted by $\mathbf{a}_f \in \mathbb{R}^{H \times W \times D}$, where $D$ is the embedding dimension. This choice preserves the model's ability to handle varying input resolutions. 

Next, we patchify our latent inputs $\mathbf{a}_f$ into a sequence of tokens $\mathbf{a}_p \in \R^{ \left(\frac{H}{P} \times \frac{W}{P}\right)\times D}$  using the same patch embedding process as in standard ViTs \cite{dosovitskiy2020image}, where $P$ denotes the patch size.  To  provide the model with information about the spatial arrangement of the patches, we add trainable spatial positional embeddings to each token as
\begin{align*}
    \mathbf{a}_{pe} = \mathbf{a}_{p}  + \text{PE}, \quad  \text{PE} \in \R^{\left(\frac{H}{P} \times \frac{W}{P}\right)  \times D}.
\end{align*}
We then process the tokens $\mathbf{a}_{pe}$ using a sequence of $L$ pre-norm  Transformer blocks \cite{vaswani2017attention, xiong2020layer},
\begin{align*}
\mathbf{z}_{0} &= \operatorname{LN}(\mathbf{a}_{pe}), \\
\mathbf{z}_{\ell}^{\prime} & =\operatorname{MSA}\left(\operatorname{LN}\left(\mathbf{z}_{\ell-1}\right)\right)+\mathbf{z}_{\ell-1}, & & \ell=1 \ldots L, \\
\mathbf{z}_{\ell}, & =\operatorname{MLP}\left(\operatorname{LN}\left(\mathbf{z}_{\ell}^{\prime}\right)\right)+\mathbf{z}_{\ell}^{\prime}, & & \ell=1 \ldots L. 
\end{align*}

\paragraph{Decoder.} The decoder module draws inspiration from the Continuous Vision Transformer (CViT) \cite{wang2024bridging} that enables the continuous evaluation of the model's outputs at any arbitrary query coordinates.
To this end, we start by creating a uniform grid $\{ \mathbf{x}_{ij}\} \subset [0, 1]^2$, for $i=1, \dots N_x$ and $j=1, \dots N_y$, where the hyperparameters $N_x, N_y$ are typically set to match the resolution of the target output functions \cite{peng2020convolutional, jiang2020local,mehta2021modulated}. For a single query point $\mathbf{x} \in \R^2$, we then perform a Nadaraya-Watson interpolation \cite{nadaraya1964estimating,watson1964smooth} over
the outputs of the FNO encoder as 
\begin{align}
\label{eq:cvit_posenc}
    \mathbf{x}' = \sum_{i=1}^{N_x} \sum_{j=1}^{N_y} w_{ij} \mathbf{a}_f^{ij},  \quad w_{ij}  =  \frac{\exp(-\beta \|\mathbf{x} -  \mathbf{x}_{ij}\|^2  ) }{ \sum_{ij} \exp(-\beta \|\mathbf{x} -  \mathbf{x}_{ij}\|^2)},
\end{align}
where $\beta >0$ is a hyperparameter that determines the locality of the interpolated features.  It is important for determining the smoothness of the interpolation function. Specifically, larger values of $\beta$ yield more localized weight distributions $w_{ij}$, resulting in a higher-frequency interpolant that captures finer-scale variations. Conversely, smaller $\beta$ values produce a smoother interpolant by averaging over a broader neighborhood of points.

We then treat the interpolated latent feature $\mathbf{x}_{0}  = \mathbf{x}' \in \R^{1 \times D}$ as queries and  the encoder output $\mathbf{z}_L$ 
as keys and values, applying  $K$ cross-attention Transformer blocks as
\begin{align*}
    \mathbf{x}^{\prime}_{k} & = \mathbf{x}_{k - 1} + \operatorname{MHA}\left(\operatorname{LN}\left(\mathbf{x}_{k - 1} 
    \right), \operatorname{LN}\left(\mathbf{z}_L
    \right), \operatorname{LN}\left(\mathbf{z}_L
    \right)\right), & & k=1 \ldots K, \\
\mathbf{x}_{k} & =\mathbf{x}^{\prime}_{k} + \operatorname{MLP}\left(\operatorname{LN}\left( \mathbf{x}^{\prime}_{k}\right)\right), & & k=1 \ldots K.
% \mathbf{y}_{K+1} &= \operatorname{LN}\left(\mathbf{y}_K
%     \right)
\end{align*}
The final output of our Micrometer model represents the strain concentration tensor $\mathbb{A} \in \R^{3 \times 3}$, flattened into a 1D vector in $\R^9$. To obtain this, we project the high-dimensional latent outputs of the model onto the physical space of the desired dimension through a small MLP.

It is worth noting that the Micrometer decoder performs  cross-attention between each individual query point and the output tokens of the encoder. This is a subtle but significant difference from prior work, such as Oformer \cite{li2022transformer} and GNOT \cite{hao2023gnot}, where model predictions from different query locations are correlated because cross-attentions are computed for all queries simultaneously. In contrast, Micrometer vectorizes the decoder module across query coordinates.   Consequently, the model predictions corresponding to different query coordinates are independent of each other, thereby allowing us to build a well-defined continuous representation.

\section{Results}
\label{sec: results}

In this section, we demonstrate the effectiveness of the proposed Micrometer model. Our evaluation consists of three main components: (a) a comparative analysis of Micrometer's performance against established neural operators and PDE surrogates; (b) an application of Micrometer to two fundamental problems in micromechanics --  computational homogenization and concurrent multiscale modeling; (c) an examination of Micrometer's transfer learning capabilities in out-of-distribution scenarios with scarce data.

\subsection{Prediction of Microscale Strain Fields}
Here we validate Micrometer's accuracy in predicting microscale strain fields across a wide range of composite material microstructures and material properties. We also compare these results against other popular neural operator surrogates using the extensive CSMBech/CMME dataset described in Section \ref{sec:data_generation}.

\paragraph{Micrometer configurations.} We consider models with various configurations, as summarized in Table \ref{tab:cvit_models}. We mostly follow the recommended hyperparameter settings in  Wang {\em et al.} \cite{wang2024bridging}, as these choices have been validated by extensive ablation studies. Specifically, unless otherwise stated, for all configurations, we employ an  FNO encoder with four layers and a cross-attention decoder with 2 layers, and set the resolution of the grid to match the dataset resolution while using $\beta=10^5$ to ensure sufficient locality of the interpolated features, thereby capturing sharp transitions in solutions.

\begin{table}[h!]
    % \small
      \renewcommand{\arraystretch}{1.2}
    \centering
    \caption{Different Micrometer  configurations considered in this work.}
    \begin{tabular}{lccccccc}
        \toprule
        \textbf{Model} & \textbf{FNO width/modes} &  \textbf{Encoder layers}  &  \textbf{Embedding dim} & \textbf{MLP width} & \textbf{Heads} & \textbf{\# Params} \\
        \midrule
        Micrometer-S & 32/32 & 4  & 256 & 256 & 8 & 20 M  \\
        Micrometer-B & 64/32 & 6  & 512 & 512 & 16 & 70 M  \\
        Micrometer-L & 64/64 & 8  & 512 & 1024 & 16 & 292 M  \\
        \bottomrule
    \end{tabular}
    \label{tab:cvit_models}
\end{table}

\paragraph{Baselines.} We select the following methods as our baselines for comparison.
\begin{itemize}
\item \textbf{Fourier Neural Operator (FNO)} \cite{li2021fourier}: An efficient framework for operator learning in the frequency domain.
\item \textbf{U-Net} \cite{ronneberger2015u}: A popular network backbone widely used for segmentation and generative modeling.
\item \textbf{Vision Transformer (ViT)} \cite{dosovitskiy2021an}: A powerful transformer-based architecture adapted for image-related tasks.
\end{itemize}

For each model, we conduct comprehensive ablation studies to optimize its respective hyperparameters, adhering to the guidelines provided in the original  publications. Detailed information on the implementation of baseline models is presented in Appendix \ref{appendix: baselines}.

We employ a unified training recipe for all experiments.  We employ the  AdamW optimizer \cite{kingma2014adam,loshchilov2017decoupled} with weight decay set to  $10^{-5}$. Our learning rate schedule includes an initial linear warm-up phase of $5,000$ steps, starting from zero and gradually increasing to  $10^{-3}$, followed by an exponential decay at a rate of $0.9$ for every $5, 000$ steps.  
To stabilize training process, we clip all gradients at a maximum norm of 1. If a loss blowup occurs, we restart training from the last saved checkpoint.

The loss function is a mean squared error (MSE) between the model predictions and the corresponding targets evaluated at randomly sampled query coordinates,
\begin{align}
\text{MSE} = \frac{1}{B Q} \sum_{i=1}^3 \sum_{j=1}^3 \sum_{k=1}^B \sum_{l=1}^Q \left| \hat{A}^{(k)}_{ij}(\mathbf{x}_l) - A^{(k)}_{ij}(\mathbf{x}_l) \right|_2^2,
\end{align}
where $A^{(k)}_{ij}(\mathbf{x}_l)$ denotes the $ij$-th variable of the $k$-th sample in the training dataset, evaluated at a query coordinate $\mathbf{x}_l$, and  $\hat{A}$ denotes the corresponding model prediction.
All models are trained for $10^5$ iterations with a batch size $B=16$.
Within each batch, we randomly sample $Q = 4,096$ query coordinates from the grid and corresponding output labels.

\paragraph{Evaluation.}
After training, we evaluate model accuracy on the test dataset using several commonly used metrics: relative $L^1$ norm, relative $L^2$ norm, and root of mean square error (RMSE),
\begin{align}
\text{Rel. } L^1 &= \frac{1}{ 9 N_{\text{test}}}  \sum_{i=1}^3 \sum_{j=1}^3 
\sum_{k=1}^{N_{\text{test}}} \frac{\| \hat{A}^{(k)}_{ij} - A^{(k)}_{ij} \|_1}{ \| A^{(k)}_{ij}\|_1}, \\
\text{Rel. } L^2 &= \frac{1}{ 9 N_{\text{test}}}  \sum_{i=1}^3 \sum_{j=1}^3 
\sum_{k=1}^{N_{\text{test}}} \frac{\| \hat{A}^{(k)}_{ij} - A^{(k)}_{ij} \|_2}{ \| A^{(k)}_{ij}\|_2}, \\
\text{RMSE} &= \sqrt{  \frac{1}{9 N_{test}} \sum_{i=1}^3 \sum_{j=1}^3  \sum_{k=1}^{N_{test}} \left\| \hat{A}^{(k)}_{ij} - A^{(k)}_{ij} \right\|_2^2},
\end{align}
where the norm is computed over the prediction at all pixel grid points, averaged over each variable of interest.

Table \ref{tab: comparison} presents the performance results of Micrometer compared to several competitive and highly optimized baselines, with comprehensive experimental results provided in Table \ref{tab:different_models}.
Our largest model (Micrometer-L) achieves the lowest error across all metrics: relative $L^1$, relative $L^2$, and RMSE. Notably, Micrometer demonstrates superior performance with fewer or comparable parameters than the other baselines.

\begin{figure}[htbp]
    \centering
    \begin{subfigure}[b]{1.0\textwidth}
        \centering
        \includegraphics[width=\textwidth]{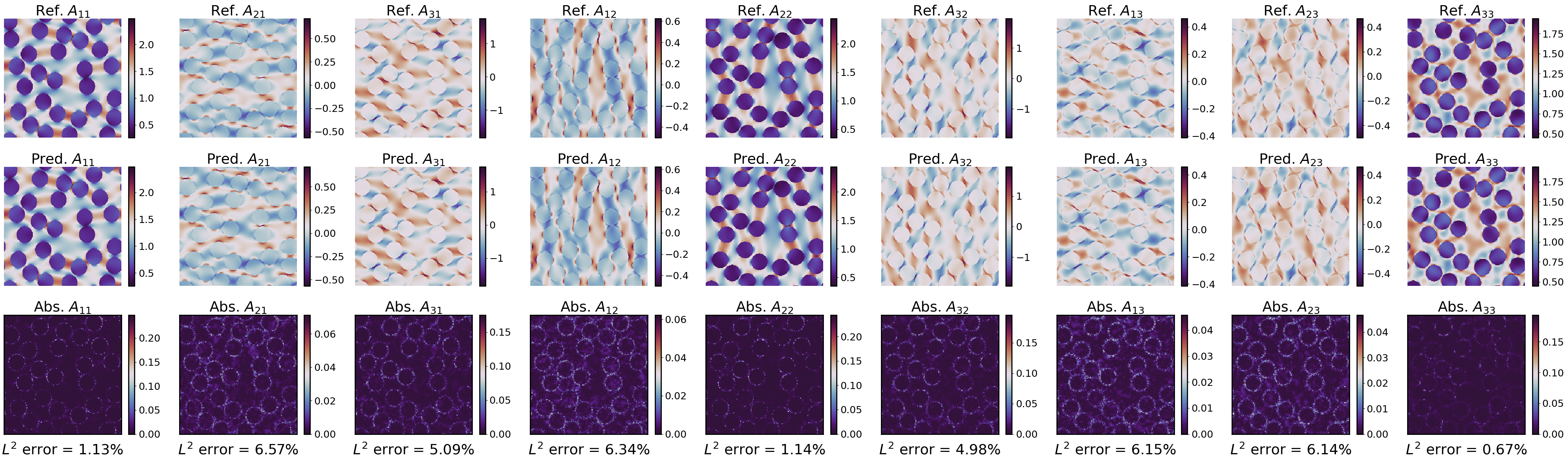}
        \caption{}
        \label{fig:pred_low_vf_ratio}
    \end{subfigure}

    \begin{subfigure}[b]{1.0\textwidth}
        \centering
        \includegraphics[width=\textwidth]{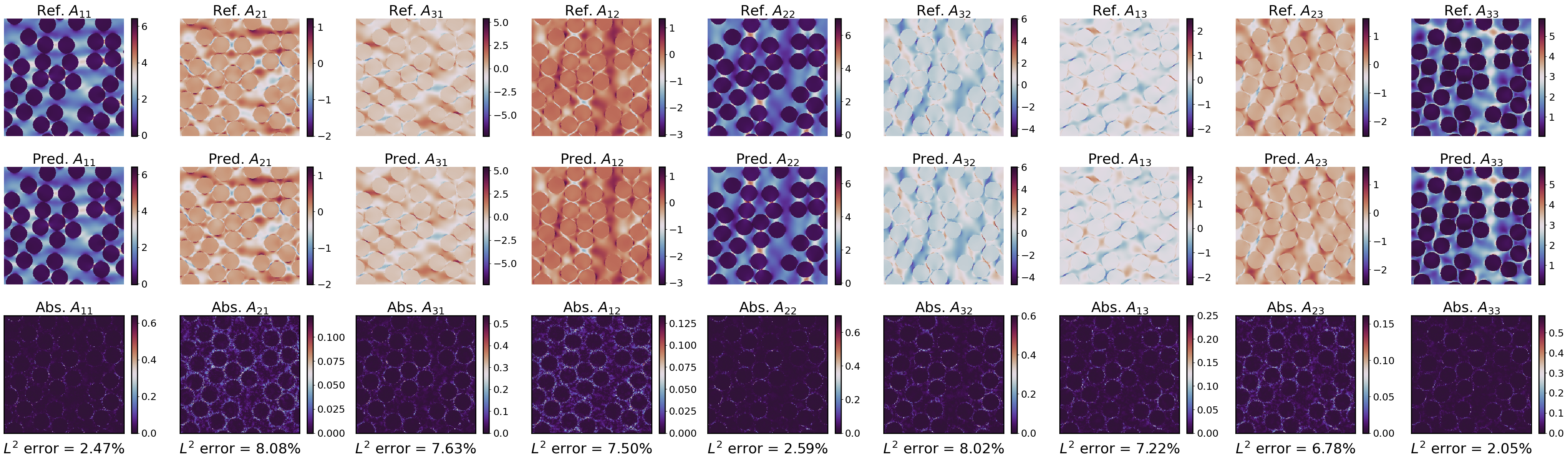}
        \caption{}
        \label{fig:pred_medium_vf_ratio}
    \end{subfigure}
 
    \begin{subfigure}[b]{1.0\textwidth}
        \centering
        \includegraphics[width=\textwidth]{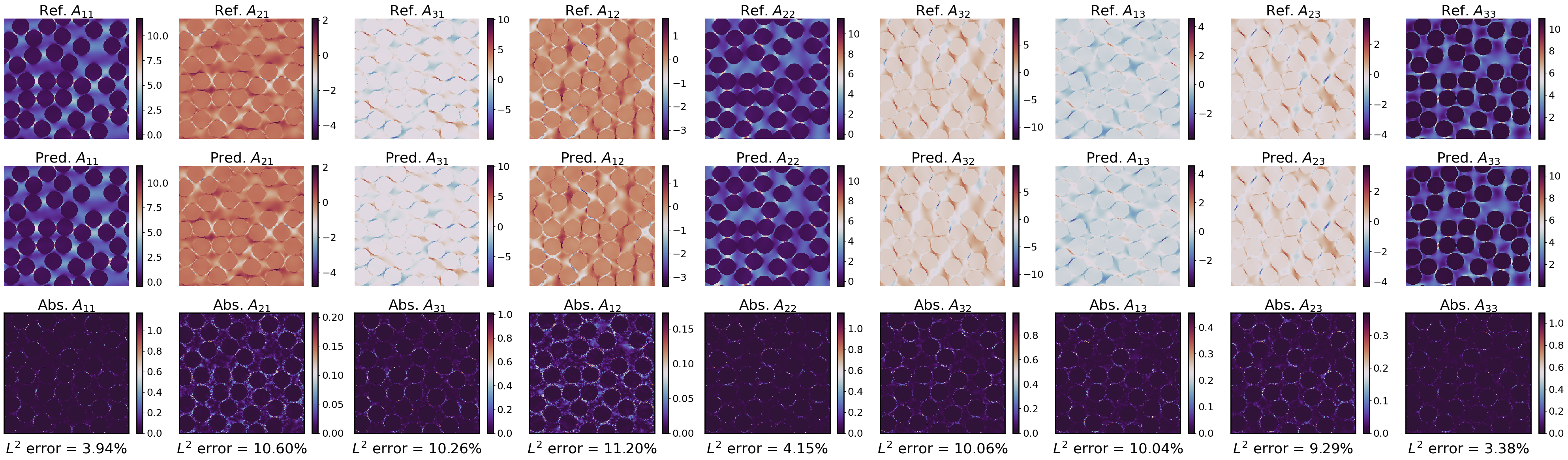}
        \caption{}
        \label{fig:pred_high_vf_ratio}
    \end{subfigure}
    
    \caption{Representative predictions by Micrometer for microstructures with varying volume fractions and Young's modulus ratios between fiber and matrix. (a) Low volume fraction and low Young's modulus ratio. (b) Medium volume fraction and medium Young's modulus ratio. (c) High volume fraction and high Young's modulus ratio.}
    \label{fig:merged_predictions}
\end{figure}

To better understand the performance of Micrometer, we visualize the test relative $L^2$ error with respect to varying volume fractions and Young's modulus ratios $E_f / E_m$ . One can see that the error increases as the volume fractions or Young's modulus ratios increase.
This trend is potentially attributed to the increased complexity of the solution under these conditions, as higher volume fractions lead to greater strain concentration in the matrix, while higher ratios result in sharper discontinuities. These observations align with the visualizations of representative predictions for low to medium volume fractions and Young's modulus ratios, as illustrated in Figure \ref{fig:merged_predictions}.

\begin{figure}
    \centering
    \includegraphics[width=0.8\linewidth]{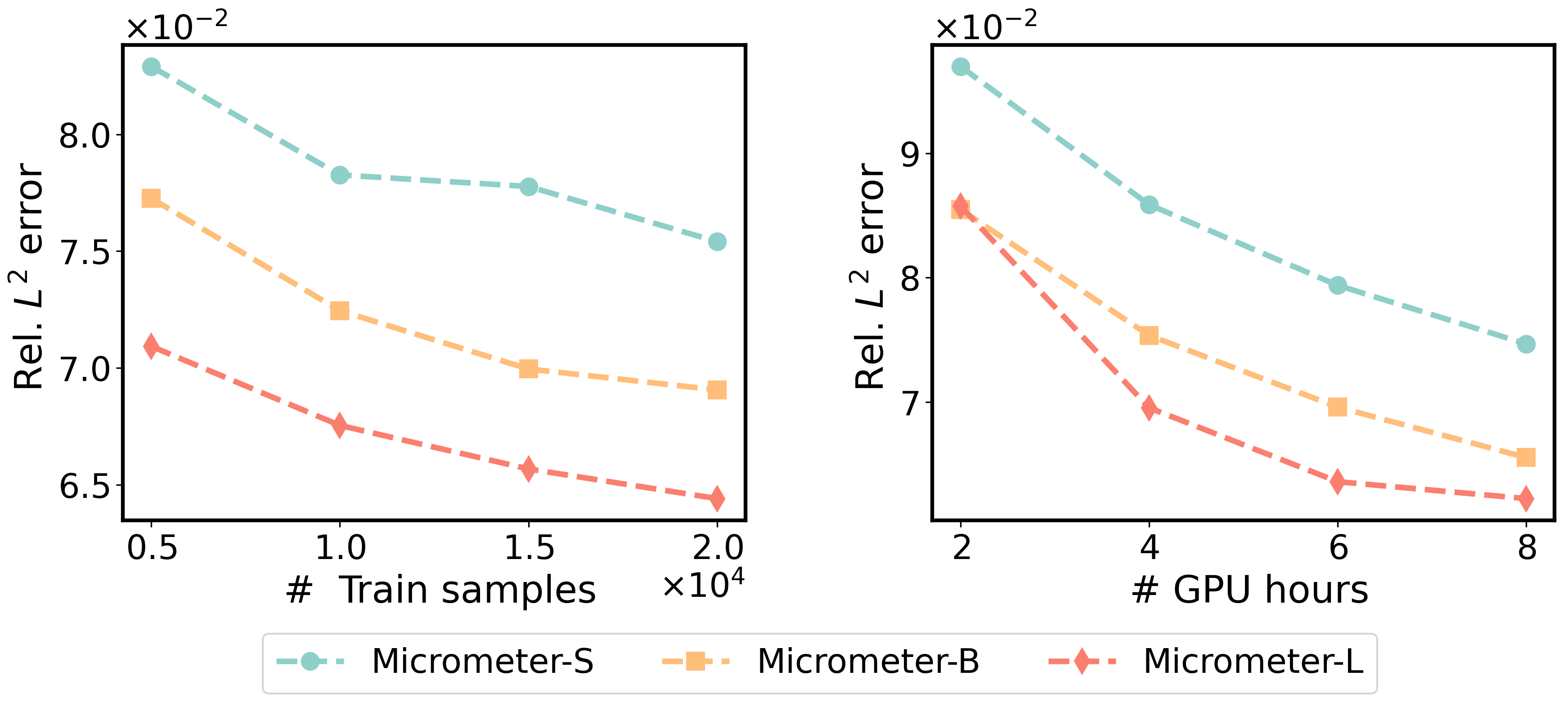}
    \caption{{\em Scaling Analysis of Micrometer:} Left: Relative $L^2$ errors for various model configurations as a function of training sample size. Right: Relative $L^2$ errors for different model configurations as a function of GPU hours. The results demonstrate the favorable scaling properties of Micrometer, with performance improvements observed for larger models, increased training samples, and additional computational resources.}
    \label{fig:scaling}
\end{figure}

Furthermore,  we conduct a scalability study of Micrometer, examining its performance with respect to the number of training samples and computational resources utilized. As shown in Figure \ref{fig:scaling}, larger models, increased training samples, and longer training times lead to improved accuracy. 
These results highlight the strong scaling performance of Micrometer, 
indicating that larger models can achieve better performance under a fixed computational budget. It appears to be in accordance with the scaling laws of transformer-based architectures observed in other domains, including natural language processing \cite{hoffmann2022training, kaplan2020scaling} and computer vision \cite{zhai2022scaling,alabdulmohsin2024getting}.

\begin{table}
\renewcommand{\arraystretch}{1.4}
    \centering
    \caption{{\em Micrometer vs. State-of-the-Art PDE Surrogate Models:} Performance of the Micrometer model against popular PDE surrogates over the test dataset. Metrics are relative $L^1$, $L^2$ errors, and root-mean-square error (RMSE). Each baseline model is optimized through hyperparameter tuning, with their best results reported here. Full  experimental results are provided in Table \ref{tab:different_models}.
    The Micrometer is available in Small, Base, and Large versions, with the Large variant achieving the best results, outperforming all baseline methods. }
  \begin{tabular}{l c c c c c}
        \toprule
        \textbf{Model} & \textbf{\# Params} & \textbf{Rel. $L^1$ error ($\downarrow$)}  & \textbf{Rel. $L^2$ error ($\downarrow$)}   & \textbf{RMSE ($\downarrow$)}\\
        \midrule
        UNet   & 124 M &  14.29 \%& 14.75  \% & 0.0752 \\
        FNO   & 268 M   &  4.79 \%   &  7.19 \%  &  0.0373   \\
        ViT    &  86 M  & 4.96 \%  & 6.86 \%   & 0.0346 \\
        \midrule
        \textbf{Micrometer-S}   & 20 M &  5.32 \%  & 7.24 \% & 0.0359  \\
        \textbf{Micrometer-B}  & 70 M  & 4.39 \%  &   6.42 \%       &    0.0337   \\
        \textbf{Micrometer-L}  & 292 M &  3.61 \%  & 5.95 \% & 0.0303  \\
        \bottomrule
    \end{tabular}
    \label{tab: comparison}
\end{table}

\begin{figure}
    \centering
    \includegraphics[width=0.8\linewidth]{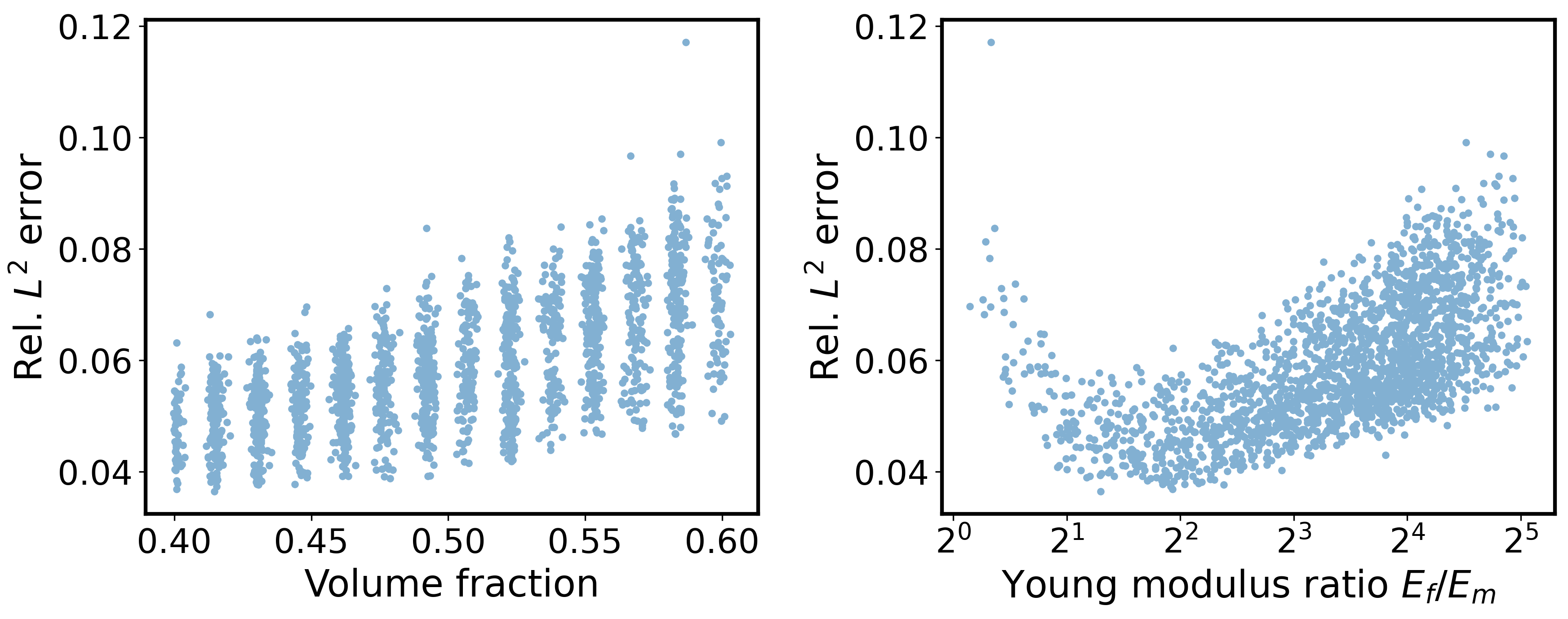}
    \caption{{\em Performance analysis of Micrometer across volume fractions and material properties}: Left: Distribution of relative $L^2$ errors with respect to volume fraction across the test dataset. Right: Distribution of relative $L^2$ errors with respect to Young's modulus ratio between fiber and matrix, evaluated over all test samples.}
    \label{fig:error_dist}
\end{figure}

% \begin{figure}[h]
%     \centering
%     \includegraphics[width=1.0\linewidth]{figures/comparisons/pred_191.png}
%     \caption{Representative prediction by Micrometer for a microstructure with low volume fraction and low Young's modulus ratio between fiber and matrix. }
%     \label{fig:pred_low_vf_raito}
% \end{figure}

% \begin{figure}
%     \centering
%     \includegraphics[width=1.0\linewidth]{figures/comparisons/pred_1066.png}
%     \caption{Representative prediction by Micrometer for a microstructure with high volume fraction and high Young's modulus ratio between fiber and matrix. }
%     \label{fig:pred_medium_vf_raito}
% \end{figure}

% \begin{figure}[h]
%     \centering
%     \includegraphics[width=1.0\linewidth]{figures/comparisons/pred_1827.png}
%     \caption{Representative prediction by Micrometer for a microstructure with high volume fraction and high Young's modulus ratio between fiber and matrix. }
%     \label{fig:pred_high_vf_raito}
% \end{figure}

\subsection{Computational Homogenization} 

\begin{table}[]
\renewcommand{\arraystretch}{1.4}
    \centering
     \caption{Comparison of computational efficiency and accuracy between traditional methods (FFT, FE-FFT) and Micrometer-based approaches for homogenization and multiscale modeling tasks. Accuracy is measured by the relative error, and computational efficiency by runtime in seconds on respective hardware and software.}
    \begin{tabular}{c|ccc}
    \toprule
        \textbf{Task} &  \textbf{Method}  & \textbf{Rel. error} & \textbf{Computational time (secs)}  \\
    \hline
     \multirow{2}{*}{Homogenization}    & FFT   &  - &  3,001 \\ 
                                        & Micrometer  & $0.07\% \pm 0.05 \%$    & 35   \\ 
    \hline
     \multirow{2}{*}{Multiscale modeling}    & FE-FFT   &  -  &  21,242\\ 
                                             & FE-Micrometer   &  $0.84 \% \pm 0.47 \% $ & 277   \\
    \bottomrule
    \end{tabular}
    \label{tab:computational_cost}
\end{table}

\begin{figure}
    \centering
    \includegraphics[width=0.4\linewidth]{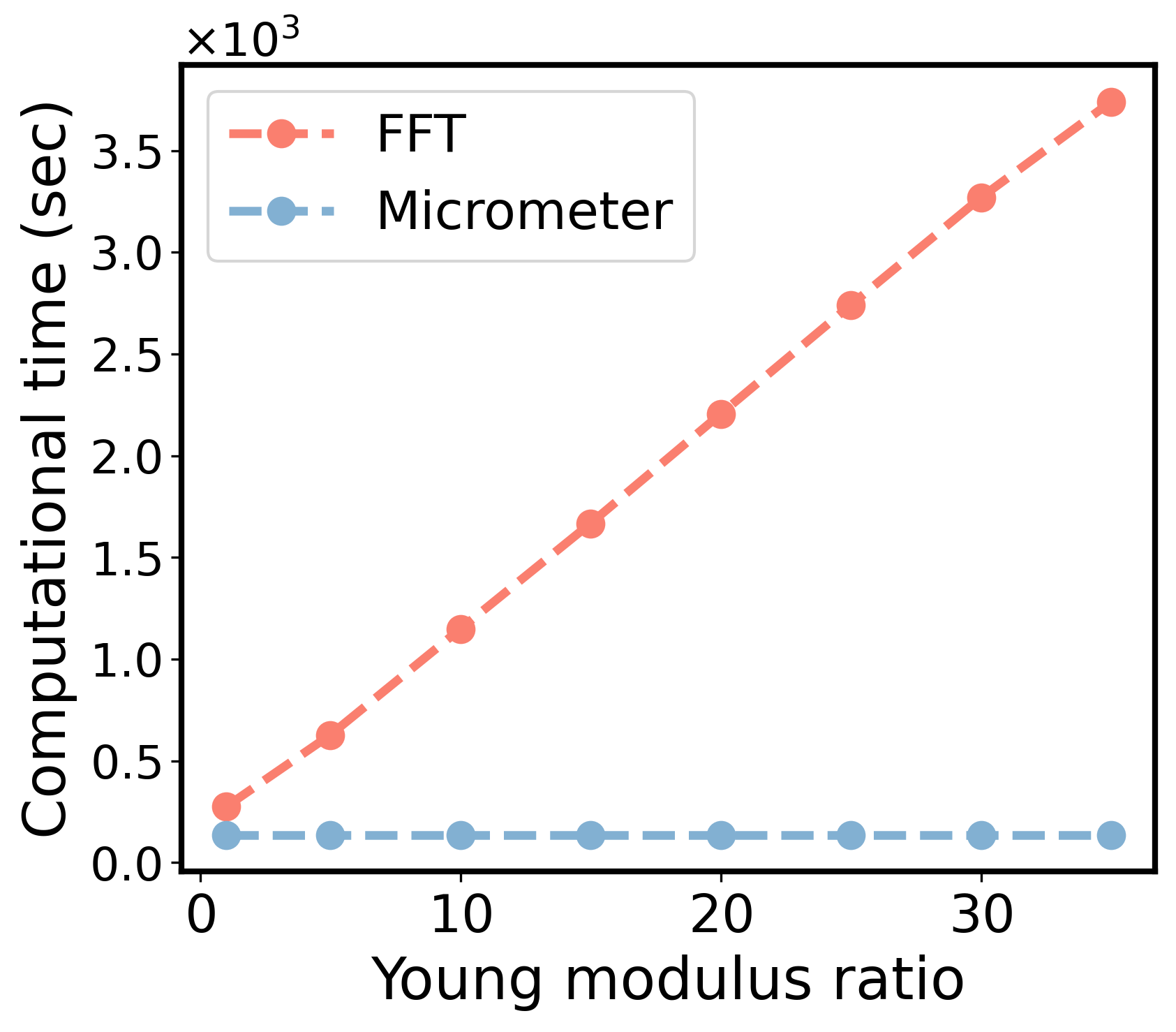}
    \caption{{\em Computational cost regarding \text{FFT based homogenization} vs. Micrometer:}
    Wall-clock time of using the conventional FFT based homogenization versus the Micrometer to predict the full field strain concentration tensor for 3,000 microstructures with volume fractions uniformly distributed between 40\% and 60\%. The FFT based homogenization was implemented by MATLAB$^\circledR$ R2024a and tested on an HP precision Z2G5 mini working station (CPU: Intel$^\circledR$ Xeon$^\circledR$ W1290 3.2 GHz 10-cores), while the Micrometer model was evaluated on NVIDIA$^\circledR$ A100 GPUs.  We acknowledge that this may not represent a unique and entirely fair comparison in terms of computational costs, as the hardware and software environments differ. This result is intended to provide a general sense of the relative runtime performance using common computational resources.}
    \label{fig:E_ratio_time}
\end{figure}

Homogenization addresses the fundamental problem of determining the effective properties of heterogeneous materials based on their microstructural characteristics.
Herein, our goal is to determine the effective material properties including Young's modulus $\bar{E}$ and Poisson ration $\bar{\nu}$ for heterogeneous RVE by using the model prediction $\mathbb{A}$ and the fourth order elastic tensor $\mathbb{C}$. Specifically, for a single RVE, we aim to obtain the homogenized fourth order elastic tensor $\bar{\mathbb{C}}$ defined by
\begin{equation}
    \bar{\mathbb{C}} = \frac{1}{T_1 T_2}\sum_{j=1}^{T2}\sum_{i=1}^{T1} \mathbb{C}(\mathbf{x}_{ij})::\mathbb{A}(\mathbf{x}_{ij}),
    \label{homogenizationrule}
\end{equation}
where $\bar{\mathbb{C}}$ is a symmetric 3 $\times$ 3 matrix in Voigt notation. Using the relation in Eq.\eqref{setup:Lametransform} and \eqref{setup:Lameconstant}, homogenized Young's modulus $\bar{E}$ and Poisson ration $\bar{\nu}$ can be expressed as:
\begin{equation}
\bar{\nu}=\frac{\bar{C}_{1111}-2\bar{C}_{1212}}{2(\bar{C}_{1111}-\bar{C}_{1212})}, \quad
\bar{E} = \bar{C}_{1212}(\frac{3\bar{C}_{1111}-4\bar{C}_{1212}}{\bar{C}_{1111}-\bar{C}_{1212}}).
\label{effectiveEnu}
\end{equation}
To demonstrate the effectiveness of Micrometer in computational homogenization for a wide range of materials, we select four representative industrial fiber reinforced composites, with their material properties detailed in Table \ref{tab:material_homo_data}. To evaluate the influence of the fiber volume fraction on effective properties, we set five groups of fiber volume fraction evenly distributed between 40\% and 60 \% and each group contains 250 RVE samples with different microstructural configurations. The homogenized properties for each group are computed by averaging Young's modulus $\bar{E}$ and the Poisson ratio $\bar{\nu}$ over these 250 samples.

\begin{figure}
    \centering
    \includegraphics[width=0.9\linewidth]{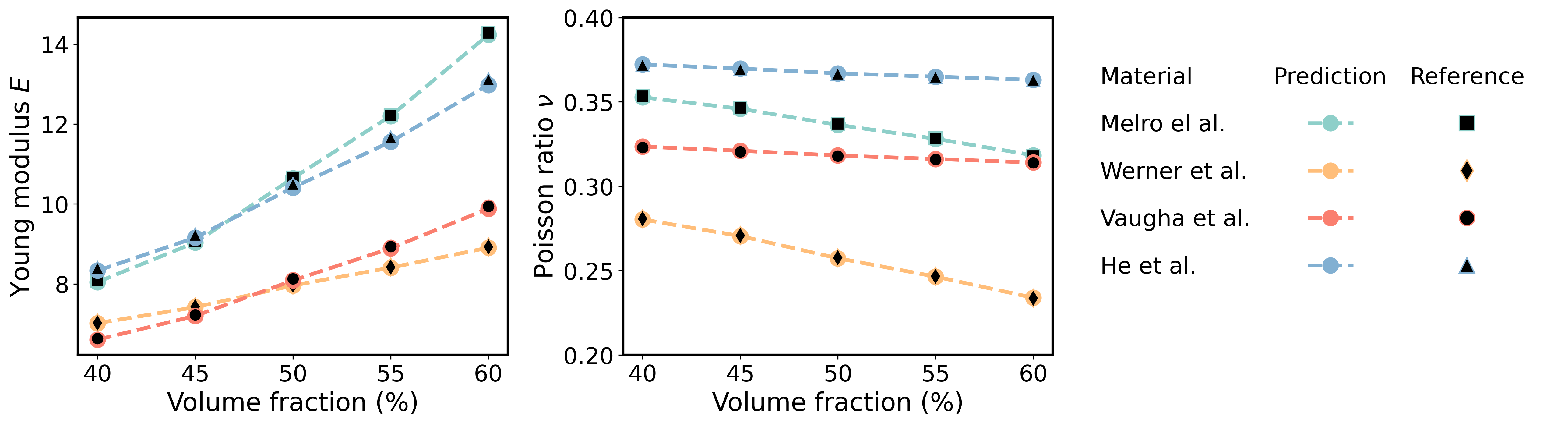}
    \caption{{\em Computational Homogenization:} Comparison of homogenized material properties obtained from conventional FFT based homogenization and Micrometer. Results are shown for four representative industrial composite materials with varying volume fractions in their representative volume elements (RVEs).}
    \label{fig: homo_error}
\end{figure}

Figure \ref{fig: homo_error} illustrates the comparison between the Micrometer's predictions and the reference values for the homogenized Young's modulus and Poisson's ratio versus volume fraction. The quantitative accuracy and the associated computational costs  are summarized in Table \ref{tab:computational_cost}. Our results show that, Micrometer achieves excellent agreement with the reference data, yielding an average relative error of less than 0.1\%, while requiring around 1\% of the computational time compared to conventional methods.
These findings highlight the accuracy and computational efficiency of the Micrometer in predicting the effective properties of heterogeneous materials. Such performance makes it a promising approach for broader applications in computational solid mechanics, particularly in the design and analysis of composite materials, where both accuracy and computational efficiency are crucial.

\begin{table}
\renewcommand{\arraystretch}{1.4}
\caption{Mechanical properties for four types of industrial composite materials. The unit of $E_m$ and $E_f$ is GPa.}
\centering
\begin{tabular}{l c c c c}
\toprule
\multicolumn{1}{c}{\textbf{Material type}} & \multicolumn{4}{c}{\textbf{Material property}} \\
\cmidrule(lr){1-1} \cmidrule(lr){2-5}
\textbf{Fiber / Matrix} & $E_f$ & $\nu_f$ & $E_m$ & $\nu_m$ \\
\midrule
Silenka E-Glass 1200tex / Bisphenol-A epoxy \cite{Melropart2} & 74.00 & 0.2000 & 3.76 & 0.39 \\
AS4 \cite{Soden1998} / 3501-6 epoxy \cite{Werner2014} & 15.00 & 0.0714 & 4.60 & 0.34 \\
HTA / 6376 \cite{Vaughan2011} & 28.00 & 0.3300 & 3.63 & 0.34 \\
T300 / TDE 86 epoxy \cite{He2020} & 40.00 & 0.3986 & 4.35 & 0.39 \\
\bottomrule
\end{tabular}

\label{tab:material_homo_data}
\end{table}

% \begin{table}
% \renewcommand{\arraystretch}{1.4}
%     \centering
%   \begin{tabular}{l c c c c c}
%         \toprule
%         \textbf{Name (fiber/matrix)} & $E_f$ & $\nu_f$ & $E_m$ & $\nu_m$\\
%         \midrule
%         Silenka E-Glass 1200tex/Bisphenol-A epoxy \cite{Melropart2} & 74.00 & 0.2000 & 3.76 & 0.39 \\
%         AS4 \cite{Soden1998} /3501-6 epoxy \cite{Werner2014} & 15.00 & 0.0714 & 4.60 & 0.34 \\
%         HTA/6376 \cite{Vaughan2011} & 28.00 & 0.3300 & 3.63 & 0.34 \\
%         T300/TDE 86 epoxy \cite{He2020} & 40.00 & 0.3986 & 4.35 & 0.39 \\
%         \bottomrule
%     \end{tabular}
%     \caption{Mechanical properties for five types of industrial composite material. The unit of $E_m$ or $E_f$ is GPa. }
%     \label{tab:material_homo_data}
% \end{table}

\subsection{Multiscale Modeling}

Concurrent multiscale modeling techniques, such as $\text{FE}^2$ and FE-FFT, aim to predict the mechanical responses of engineering structures at both macroscale and microscale levels. By incorporating detailed microstructural morphologies and material properties, these approaches circumvent the need for calibrating complex macroscale constitutive equations. Broadly speaking,  through considering the material heterogeneity at the microscale, RVEs are employed as surrogates for phenomenological macroscopic data or constitutive laws, serving as intermediaries to transmit microscale information to the macroscale.

In this section, we present a novel concurrent multiscale framework called FE-Micrometer for predicting the mechanical responses of 2D fiber-reinforced composite plates. This framework integrates the finite element method (FEM) to solve the macroscale boundary value problem (MBVP) with Micrometer for resolving the microscale boundary value problem (mBVP) within the RVE, which is associated with a macroscale solid element. For the macroscale FEM, we employ plane strain quadrilateral elements with reduced (one-point) integration scheme (CPE4R). The overall algorithm for FE-Micrometer is outlined in Algorithm \ref{algo:FE_micrometer}, see Appendix \ref{appendix: fft}.

To demonstrate the efficacy of our approach, we model a macroscale composite plate in 2D, comprising 40 × 75 RVEs in the horizontal and vertical directions, respectively, resulting in 3,000 RVEs in total.
These RVEs are divided into 20 groups with fiber volume fraction uniformly distributed ranging from 40\% to 60\% and randomly distributed in spatial coordinates within the plate. To consider the uncertainty of material properties across macroscale during the manufacture process, we introduce a spatially correlated Gaussian random field (GRF) for the Young's modulus of both matrix and fiber in each RVE, by using the Karhunen–Loève (K-L) expansion:
\begin{equation}
E_{\square}(\mathbf{X}_M,\eta) = \hat{E}_{\square} + \sum_{k=1}^{N_{kl}}\sqrt{\zeta_k}\gamma_k(\mathbf{X}_M)\rho_k(\eta),
\label{klexpansion}
\end{equation}
where $\mathbf{X}_M$  denotes a macroscale material point corresponding to a single RVE, and  $\square$ represents either the fiber or matrix material domain.  In particular, $E_{\square}(\mathbf{X}_M,\eta)$ and $\hat{E}_{\square}$ represent Young's modulus at $\mathbf{X}_M$ and its global average, respectively.  $N_{kl}$ represents the number of K-L expansion modes or truncated level. $\zeta_k$ and $\gamma_k(\mathbf{X}_M)$ denote the $k$-$\text{th}$ largest eigenvalue and associated normalized eigenfunction of the covariance function, respectively. $\rho_k(\eta)$ denotes the independent standard uncorrelated random variables.

\begin{figure}
    \centering
    \includegraphics[width=0.9\linewidth]{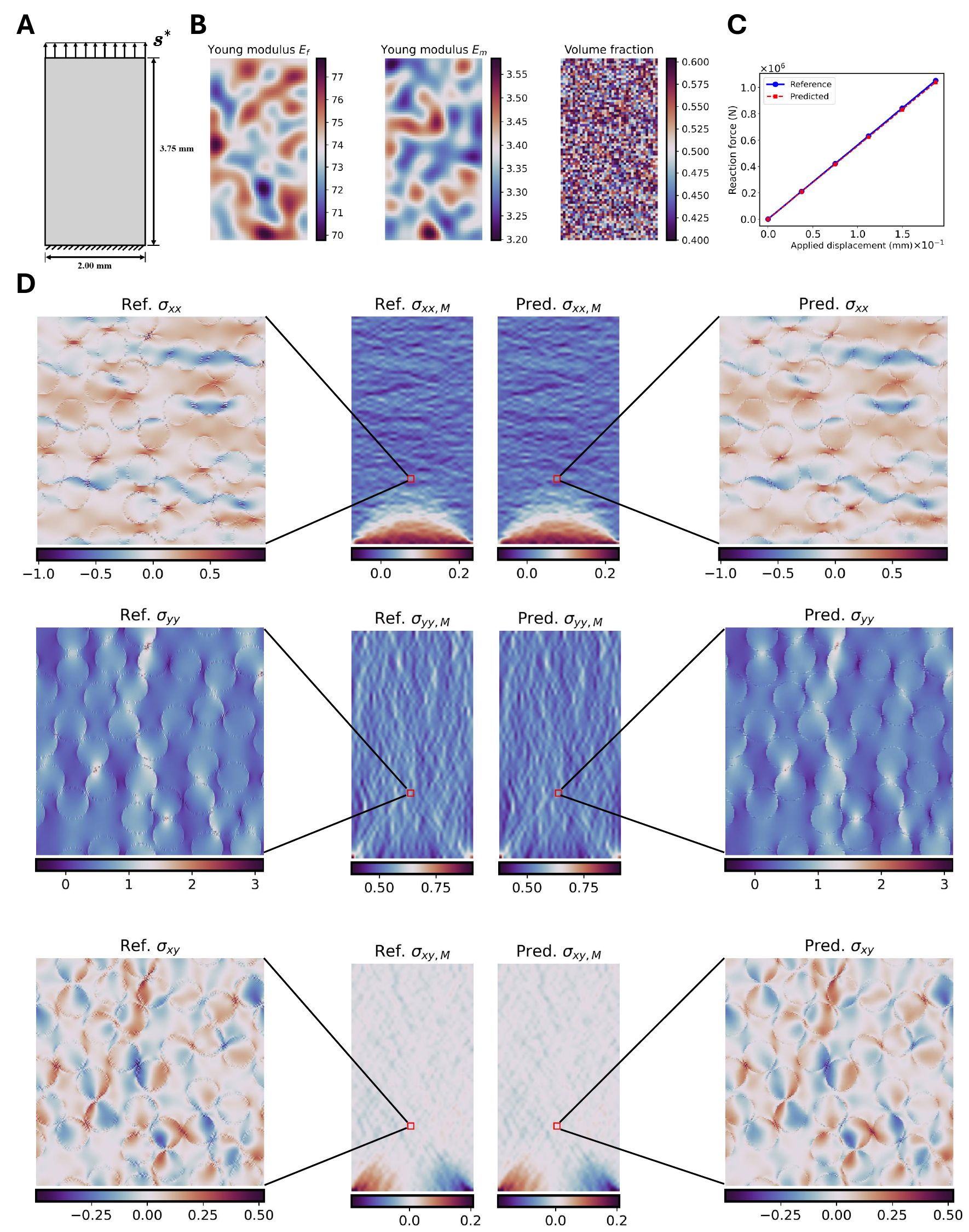}
    \caption{{\em Concurrent Mutiscale Modeling:} (A) Geometric setup and boundary conditions of the macroscale plate: fixed support at the bottom edge and uniaxial tensile load applied to the top edge.  (B) Distributions of material properties and volume fraction across the macroscale plate. (C) Comparison of reaction forces with respect to applied displacement, obtained from FE-FFT and FE-Micrometer, respectively. (D) Comparison of ultimated macroscale stress fields predicted by FE-FFT and FE-Micrometer, alongside microscale stress fields for a representative volume element (RVE) in the plate. Note that the microstress and macrostress are denoted by $\mathbf{\sigma}_{\square}$ and $\mathbf{\sigma}_{\square,M}$, respectively.}
    \label{fig:multiscale}
\end{figure}

To model a high Young's modulus ratio between fiber and matrix ($\hat{E}_f/\hat{E}_m$), we set $\hat{E}_f$ and $\hat{E}_m$ to 74 GPa and 3.35 GPa, respectively, with standard deviations of 2 GPa and 0.1 GPa. The correlation length scale is set as 0.1 mm. For macroscale loading conditions, as depicted in Panel A of Figure \ref{fig:multiscale},  we impose Dirichlet boundary conditions on the composite plate: a vertical displacement ($\boldsymbol{s}^{*} = 0.1875 \text{mm}$) is applied to the top edge and the bottom edge is fixed in both horizontal and vertical directions.  All other edges remain stress-free. We consider a quasi-static loading scenario without inertial effects, with the total loading history $\boldsymbol{s}^{}$ divided linearly into five steps.

Figure \ref{fig:multiscale} presents a comprehensive summary of our results. Panel C compares the vertical reaction force versus the applied displacements at different loading steps for both FE-FFT and FE-Micrometer. It can be observed that our predictions closely align with the reference data, with a relative error of 1.03\%.  In Panel D, we plot both the predicted and reference macrostress, along with visualizations of microstress for a RVE. 
Additional results corresponding to plates with low and medium ratios of Young's modulus are shown in Figures \ref{fig:macro_pred_low} and \ref{fig:macro_pred_medium}, respectively, which demonstrates good agreement between the predictions and the corresponding ground truth across various material property ranges.

Furthermore, Table \ref{tab:computational_cost} demonstrates that the FE-Micrometer method achieves a speedup of two orders of magnitude compared to the conventional FE-FFT approach. Such significant acceleration can  be attributed to  Micrometer's efficient handling of microscale simulations, which dominate the total computational cost. As illustrated in Figure \ref{fig:E_ratio_time}, we  compares the time required to generate 3,000 solutions using FFT based homogenization and Micrometer. It can be observed that the computational time for FFT based homogenization scales linearly with increasing Young's modulus ratio $E_f / E_m$, while Micrometer's performance remains constant, as it only involves a neural network evaluation. The results strongly suggest that we can obtain substantial acceleration from FE-Micrometer, particularly for high Young's modulus ratios. In conclusion, the significant reduction in computational cost, combined with high predictive accuracy, highlights the FE-Micrometer method's potential as an efficient and precise tool for multiscale material modeling, with broad applications across various engineering domains.

\subsection{Transfer Learning and Adaptability}

Building upon the  efficacy of Micrometer in previous applications, this section investigates its adaptability through fine-tuning. Our goal is to highlight the model's capacity for transfer learning
to new, specific tasks with minimal additional data and training. To this end, we consider two cases as follows.
 
\paragraph{Case I.} This case validates our model's adaptability to fiber-reinforced composite RVEs with different microstructure parameter settings distinct from those used in training. Specifically, we set  the fiber radius $R_d$ to 5$\mu \text{m}$ and the volume fraction range $\textit{Vof}$ to 62.5\% - 65\%. The discretization resolution, RVE size, standard deviation of fiber radius, and material properties remain the same as those in Table \ref{tab:dataset}. Following Algorithm \ref{algo:datageneration}, we generate 1,000 training and 100 test RVE samples with diverse microstructural configurations, along with corresponding input and output datasets governed by these new settings.

\paragraph{Case II.} This case examines our model's adaptability to RVEs with microstructural morphologies fundamentally different from fiber-reinforced composites. We focus on RVEs featuring microscale stochastic architected materials, specifically cellular materials shaped in spinodal topologies. Such materials find wide application in both industrial and biomedical fields, including energy absorption \cite{Annasmall2019}, human bone implants \cite{JanPNAS2022}, and impact-resilient structures \cite{Portela2021}, among others.
The spinodal RVEs are generated by solving the time-dependent Cahn-Hilliard (CH) equation using a spectral FFT method, with mathematical and computational foundations described by Chen and Shen \cite{CLQ1998}. We set the RVE size to $50 \mu \text{m} \times 50 \mu \text{m}$, with a $256 \times 256$ discretization resolution. The RVE domain is divided into hard and soft phases based on the solution field of the Cahn-Hilliard equation at each pixel, with each phase occupying 50\% of the volume. Pixels are labeled as $\textbf{0}$ (soft phase) if their concentration exceeds a threshold value of 0.6, and $\textbf{1}$ (hard phase) otherwise.
We generate 1,000 training and 100 test spinodal RVE samples with varying configurations. Material properties are assigned following \textbf{Step 1} in Algorithm \ref{algo:datageneration} (see Appendix \ref{appendix: fft}), with the matrix and fiber domains replaced by soft and hard phases, respectively. The calculation of the fourth-order elastic tensor and strain concentration tensor for all spinodal RVEs follows \textbf{Steps 2} and \textbf{3} in Algorithm \ref{algo:datageneration} (see Appendix \ref{appendix: fft}), respectively.

\begin{table}[htbp]
\centering
\caption{Summary of the dataset parameters for transfer learning tasks. }
\label{tab:dataset_comparison}
\renewcommand{\arraystretch}{1.2}
\begin{tabular}{lcc}
\toprule
\textbf{Dataset Size} & \textbf{Case I} & \textbf{Case II} \\
\midrule
 \# Training Samples & \multicolumn{2}{c}{1,000} \\
\# Test Samples & \multicolumn{2}{c}{100} \\
\midrule
\multicolumn{2}{l}{\textbf{RVE Characteristics}} \\
\midrule
Discretization resolution & \multicolumn{2}{c}{$256 \times 256$} \\
Fiber volume Fraction $\textit{Vof}$ & 62.5\% & N/A \\
Fiber radius $R_d$ $(\mu \text{m})$ & 5 & N/A \\
Fiber radius standard deviation $\textit{Std}$ & 1\% & N/A \\
\midrule
\multicolumn{2}{l}{\textbf{Material Properties}} \\
\midrule
Fiber Young’s modulus $ E_f$ (GPa)& \multicolumn{2}{c}{$5 \text{-} 85$} \\
Matrix Young’s modulus $E_m$ (GPa) & \multicolumn{2}{c}{$2.5 \text{-} 5$} \\
Fiber Poisson’s ratio $\nu_f$ & \multicolumn{2}{c}{$0.05 \text{-} 0.45$} \\
Matrix Poisson’s ratio $\nu_m$ & \multicolumn{2}{c}{$0.3 \text{-} 0.4$} \\
\bottomrule
\end{tabular}
\end{table}

\begin{figure}
    \centering
    \includegraphics[width=0.7\linewidth]{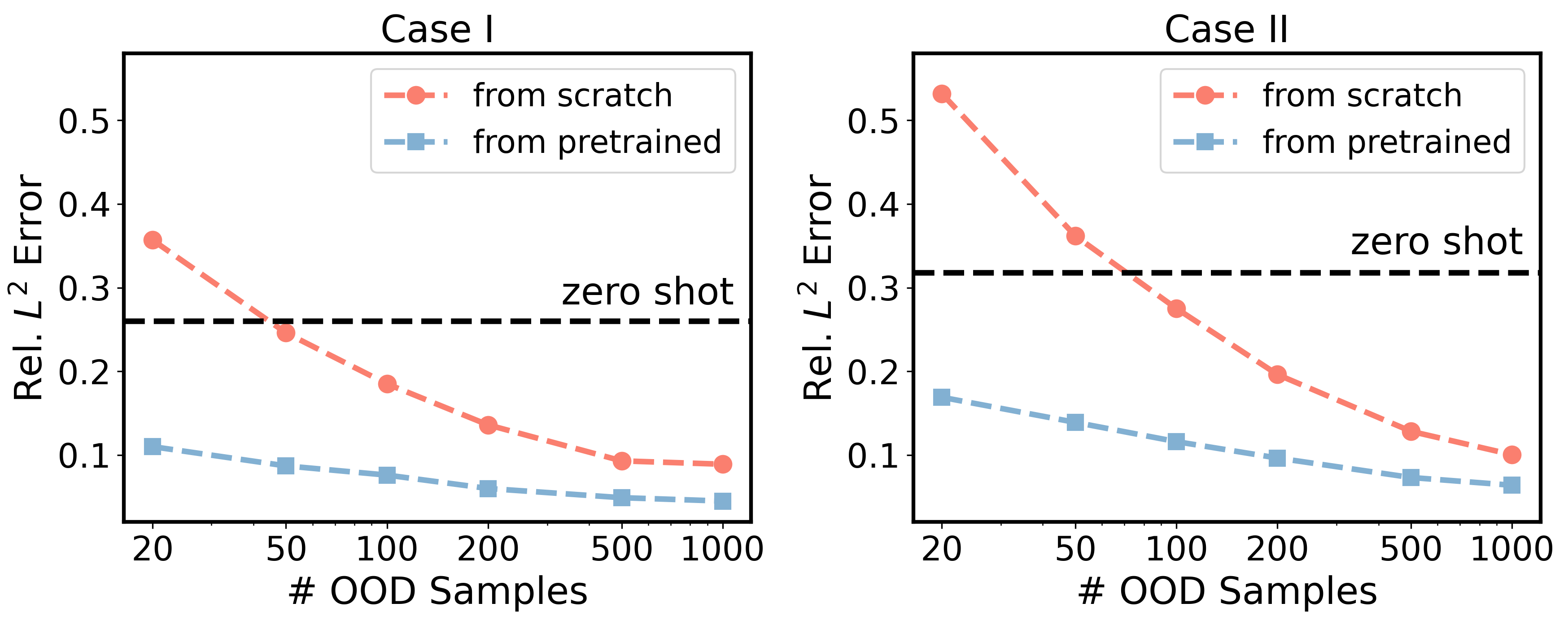}
    \caption{{\em Finetuning performance of Micrometer:} Relative $L^2$ errors for models trained from scratch versus finetuned from pretrained weights, across varying numbers of training samples. The black dashed line represents zero-shot evaluation. Results highlight Micrometer's efficient adaptation to new tasks, achieving high accuracy (below 10\% error) with fewer than 100 samples through finetuning. }
    \label{fig: tl_error}
\end{figure}

\begin{figure}[h]
    \centering
    \begin{subfigure}[b]{1.0\textwidth}
        \centering
        \includegraphics[width=\textwidth]{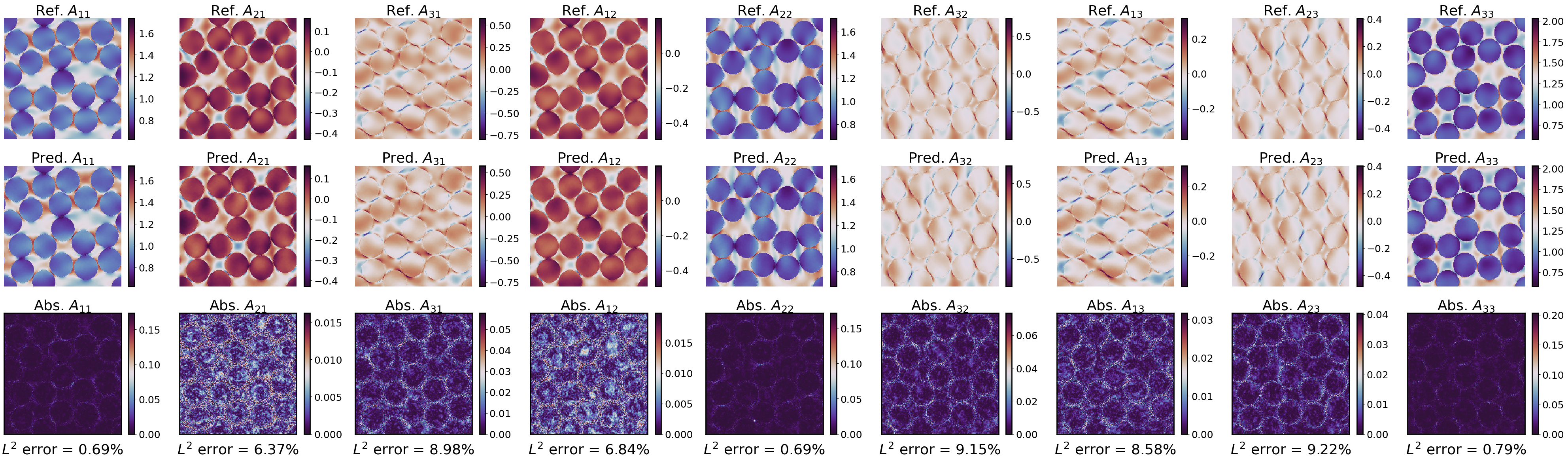}
        \caption{}
        \label{fig:tl_case1_pred_low_vf_ratio}
    \end{subfigure}

    \begin{subfigure}[b]{1.0\textwidth}
        \centering
        \includegraphics[width=\textwidth]{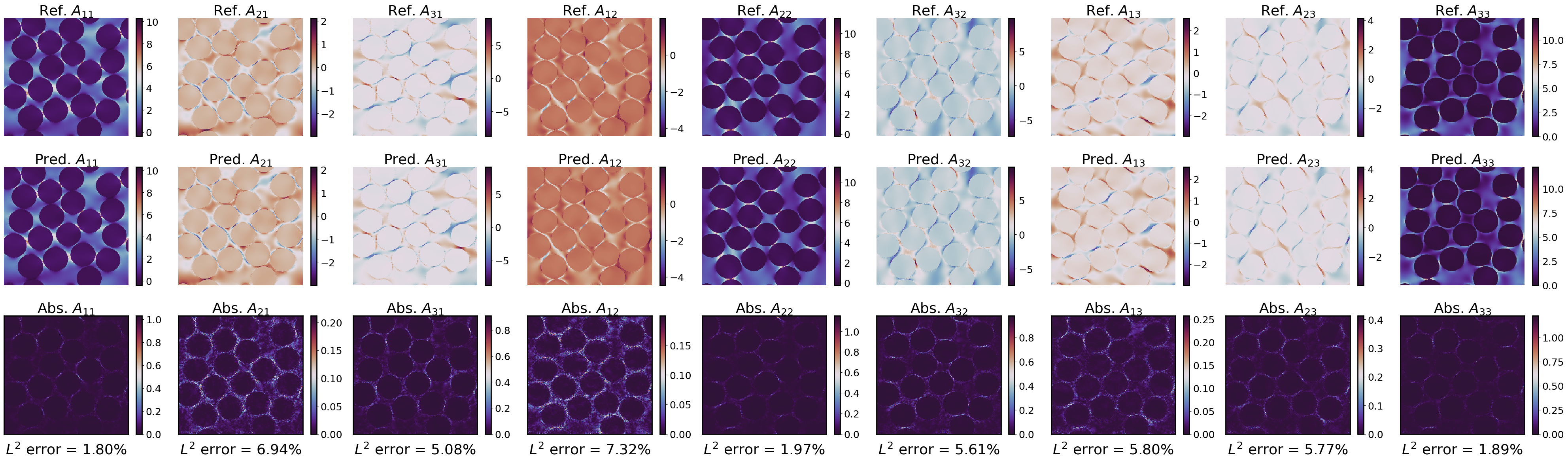}
        \caption{}
        \label{fig:tl_case1_pred_medium_vf_ratio}
    \end{subfigure}
 
    \begin{subfigure}[b]{1.0\textwidth}
        \centering
        \includegraphics[width=\textwidth]{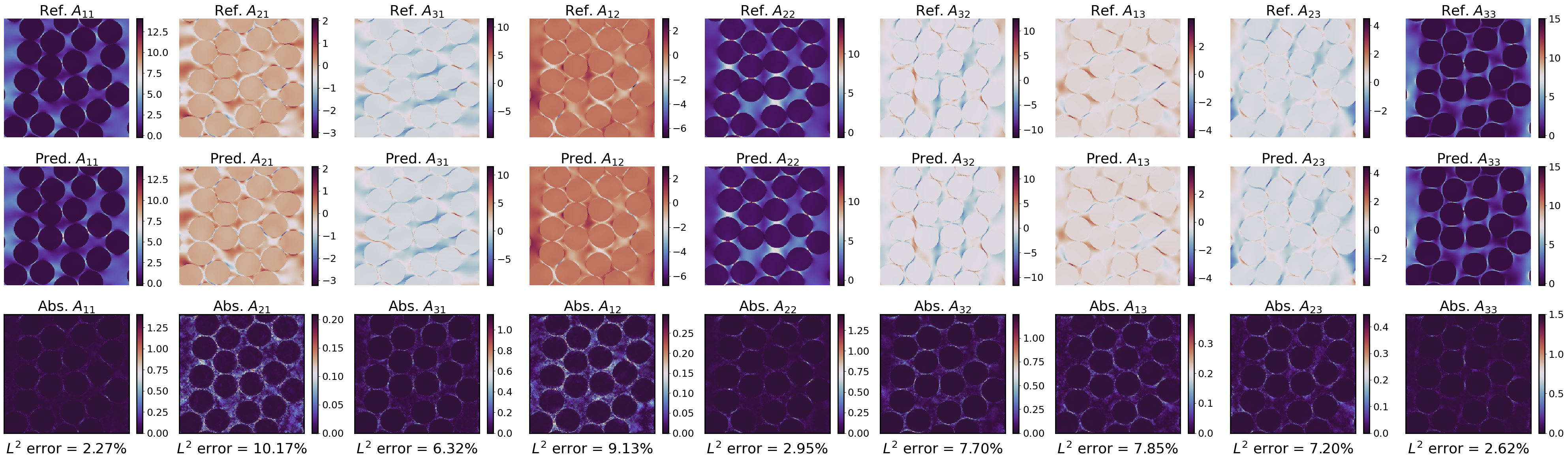}
        \caption{}
        \label{fig:tl_case1_pred_high_vf_ratio}
    \end{subfigure}
    
    \caption{{\em Transfer Learning (Case I):} Representative predictions by Micrometer for microstructures with varying volume fractions and Young's modulus ratios between fiber and matrix. (a) Low Young's modulus ratio. (b) medium Young's modulus ratio. (c) High Young's modulus ratio.}
    \label{fig:tl_case1_preds}
\end{figure}

\begin{figure}[h]
    \centering
    \begin{subfigure}[b]{1.0\textwidth}
        \centering
        \includegraphics[width=\textwidth]{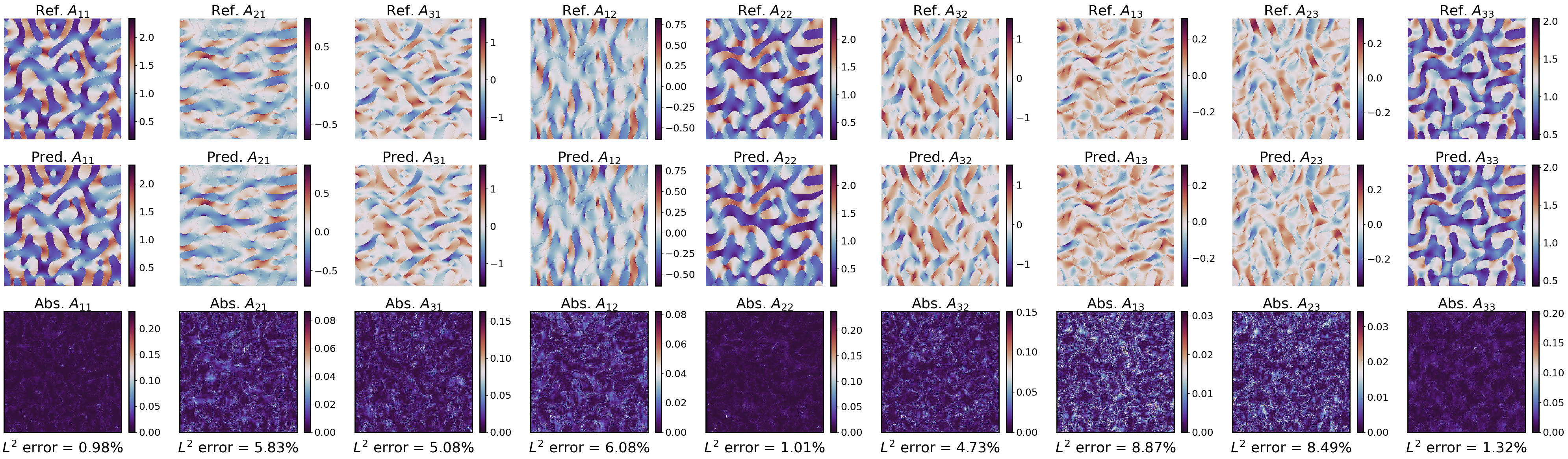}
        \caption{}
        \label{fig:tl_case2_pred_low_vf_ratio}
    \end{subfigure}

    \begin{subfigure}[b]{1.0\textwidth}
        \centering
        \includegraphics[width=\textwidth]{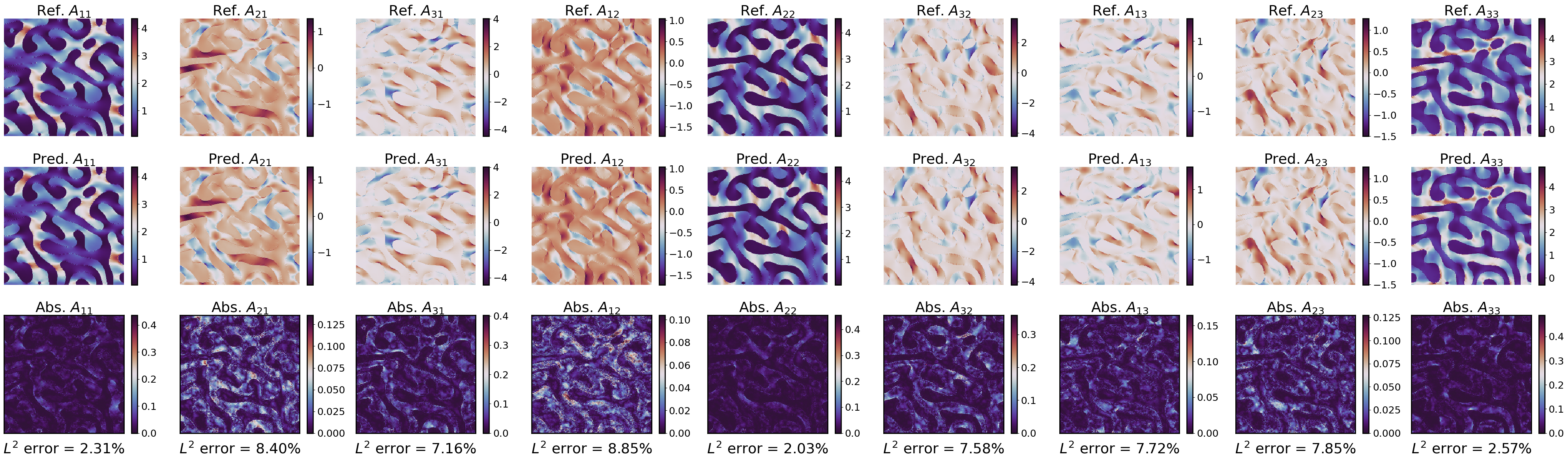}
        \caption{}
        \label{fig:tl_case2_pred_medium_vf_ratio}
    \end{subfigure}
 
    \begin{subfigure}[b]{1.0\textwidth}
        \centering
        \includegraphics[width=\textwidth]{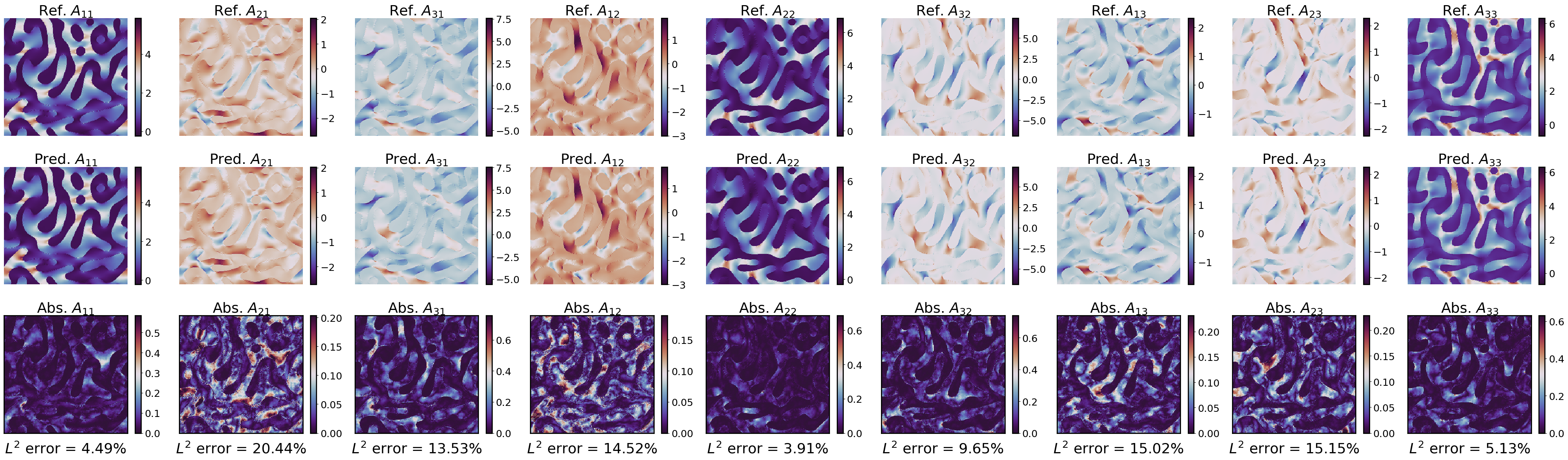}
        \caption{}
        \label{fig:tl_case2_pred_high_vf_ratio}
    \end{subfigure}
    
    \caption{{\em Transfer Learning (Case II):} Representative predictions by Micrometer for microstructures with varying volume fractions and Young's modulus ratios between fiber and matrix. (a) Low Young's modulus ratio. (b) Medium Young's modulus ratio. (c) High Young's modulus ratio.}
    \label{fig:tl_case2_preds}
\end{figure}

Figure \ref{fig: tl_error} visualizes the final relative $L^2$ error obtained after fine-tuning the pretrained Micrometer model compared to training from scratch, across varying numbers of out-of-distribution (OOD) samples.  In both Case I and Case II, fine-tuning the pretrained model consistently yields significantly lower errors, especially when the amount of OOD data is limited.
We also remark that the accuracy of zero shot evaluation for Case II is worse than that of Case I, suggesting that the Case II dataset distribution diverges further from the original training data distribution.  As a result, when training from scratch with limited data, we observe reduced  accuracy in Case II.
Moreover, Figures \ref{fig:tl_case1_preds} and \ref{fig:tl_case2_preds} depict some representative predictions from the fine-tuned model, demonstrating strong agreement with the corresponding ground truth.
Therefore, we may conclude that,  by leveraging the rich, generalizable representations learned during pretraining, our model can be efficiently adapted to specific downstream applications, while  maintaining high performance and substantially reducing the computational and data overhead typically required for training from scratch.

% \begin{figure}
%     \centering
%     \includegraphics[width=1.0\linewidth]{figures/tl/tl_vf_error.png}
%     \caption{{\em Transfer Learning case I:} Representative prediction by Micrometer for a microstructure with a volume fraction of 62.5\% and larger fiber radius. }
%     \label{fig:tl_vf_error}
% \end{figure}

% \begin{figure}
%     \centering
%     \includegraphics[width=1.0\linewidth]{figures/tl/tl_ch_error.png}
%     \caption{{\em Transfer Learning Case II:} Representative prediction by Micrometer for a microstructure generated using the Cahn-Hilliard equation. }
%     \label{fig:tl_ch_error}
% \end{figure}

\section{Discussion}
\label{sec: discussion}

In this work, we introduce Micrometer, a deep learning model for predicting the mechanical response of heterogeneous materials. This model achieves both high accuracy and computational efficiency. Our approach begins with the formulation of a novel operator learning problem, focusing on learning the solution operator of the parametric Lippmann-Schwinger equation. Specifically, we learn the mapping from the fourth-order elastic tensor to the strain concentration tensor. This formulation enables efficient evaluation of mechanical responses across a diverse range of microstructural configurations and material properties, under any external loading conditions. 

To address the scarcity of high-quality data in this field, we have generated and released CSMBench/CMME, the first large-scale, comprehensive, high-fidelity, and high-resolution dataset specifically focused on linear elasticity. Leveraging this comprehensive  dataset, we demonstrate that Micrometer outperforms several popular PDE surrogates, exhibiting strong scalability with respect to data size, model parameters, and computational resources. Moreover, we validated Micrometer's effectiveness through applications in computational homogenization and multiscale modeling, where it achieves accuracy comparable to traditional numerical methods while reducing computational time by up to two orders of magnitude. Finally, we have showcased Micrometer's adaptability through transfer learning experiments on new materials with limited data, highlighting its potential to tackle diverse scenarios in computational solid mechanics with minimal additional training.

While Micrometer represents a significant advancement in computational micromechanics, there remain several avenues for improvement and expansion. Our future work will focus on three key areas. First, we aim to extend the model to three-dimensional representations, which will enable more accurate modeling of complex material structures and behaviors, thus broadening its applicability to real-world materials and components. Second, we plan to move beyond linear elasticity by incorporating more sophisticated material behaviors such as hyperelasticity, plasticity, and fracture mechanics into the Micrometer framework. This expansion will significantly enhance the model's capability to simulate a wider range of mechanical responses under various conditions. Third, we will continue to develop and curate high-quality datasets that capture these complex phenomena, while simultaneously advancing the model architecture to handle non-linear and time-dependent mechanical responses. It is important to acknowledge that these more complex settings often come with substantially increased computational costs. However, by leveraging our accurate pre-trained model as a starting point, we can potentially accelerate future research and mitigate the computational burden associated with exploring these intricate material behaviors. This approach not only promises to extend the capabilities of Micrometer but also to make the exploration of complex material physics more computationally feasible.

Another exciting avenue  for future research is the integration of physics-informed machine learning techniques \cite{karniadakis2021physics,wang2023expert}.  This approach could substantially enhance Micrometer's predictive accuracy and data efficiency by embedding fundamental physical laws directly into the model training process. Such a methodology could be particularly beneficial  for simulating complex phenomena like crack propagation or phase transformations, where adherence to physical principles is crucial and data generation is computationally expensive. 
By integrating data-driven learning with physics-based modeling,  we anticipate that these future developments will greatly improve Micrometer's versatility and applicability, potentially accelerating materials discovery, optimizing structural designs, facilitating more efficient and accurate multiscale simulations in materials science and engineering.

% We anticipate that these future developments will significantly enhance Micrometer's versatility and applicability, potentially revolutionizing approaches to materials science and engineering. By bridging the gap between data-driven learning and physics-based modeling, Micrometer has the potential to accelerate materials discovery, optimize structural designs, and enable more efficient and accurate multiscale simulations across various industries.

% By bridging the gap between data-driven learning and physics-based modeling,  we anticipate that these future developments will significantly enhance Micrometer's versatility and applicability, potentially revolutionizing approaches to materials science and engineering in accelerating materials discovery, optimizing structural designs, facilitating more efficient and accurate multiscale simulations across various industries.

\section*{Data and Code Availability}
Code and data will be made publicly available at: \url{https://github.com/sifanexisted/micrometer}.

\section*{Acknowledgements}
We would like to acknowledge support from the US Department of Energy under the Advanced Scientific Computing Research program (grant DE-SC0024563). We also thank the developers of the software that enabled our research, including JAX \cite{jax2018github}, Matplotlib \cite{hunter2007matplotlib}, and NumPy \cite{harris2020array}.

% acknowledge Polaris
% acknowledge People for discussion

%Bibliography

\bibliographystyle{unsrt}  
\bibliography{references}

\appendix

\clearpage

\section{Nomenclature}

Table \ref{tab:notation} summarizes the main symbols and notations used in this work. The fourth-order and second-order tensors are represented in \textit{Voigt} notation. Accordingly, second-order tensors are expressed in vector form for computational convenience.

\begin{table}[!ht]
\centering
\small
\renewcommand{\arraystretch}{1.2}
\caption{Summary of the  main symbols and notation used in this work.}
\begin{tabular}
{|>{\centering\arraybackslash}m{0.2\textwidth}|>{\arraybackslash}m{0.7\textwidth}|}
\hline
\rowcolor{gray!30}
\textbf{Notation} & \textbf{Description} \\ \hline
\multicolumn{2}{|>{\columncolor{gray!15}}c|}{\textbf{Micromechanics}} \\
\hline
RVE & Representative volume element \\
PBC & Periodic boundary condition \\
FRP & Fiber reinforced composite \\
$:$ & Tensor contraction between fourth-order tensor and second-order tensor \\
$::$ & Tensor contraction between two fourth-order tensors \\
$*$ & Convolution operator \\
$\Omega \in \mathbb{R}^{2}$ & RVE domain in 2D\\
$\mathbf{x} \in \Omega$ & Spatial coordinate of a microscale material point inside RVE domain $\Omega$ \\
$\bm{\xi}$  & Frequency vector in Fourier space corresponding to $\mathbf{x}$ in Euclidean space\\
$\boldsymbol{s},\boldsymbol{\tilde{s}},\boldsymbol{\bar{s}}$ & Absolute, fluctuation and average displacement vector\\
$\boldsymbol{\varepsilon},\boldsymbol{\tilde{\varepsilon}},\boldsymbol{\bar{\varepsilon}}$ & Absolute, fluctuation and average strain vector\\ 
$\boldsymbol{\sigma}$ & Stress vector\\
$\boldsymbol{\tau}$ & Polarization stress vector\\ 
$\mathbb{C}$ & Fourth-order elastic tensor \\
$\mathbb{A}$ & Strain concentration tensor \\
$\mathbb{G}^{(0)}$, $\mathbb{\hat{G}}^{(0)}$ & Green's function of Lippmann-Schwinger equation in Eucliean and Fourier space \\
$E$, $\nu$ & A pair of Young's modulus and Poisson ratio\\
$\lambda$, $\mu$ & A pair of Lame constants\\
$T$ & RVE discretization resolution\\
$h$ & Pixel size in discretization of RVE\\
$\chi$ &  Characteristic function determines the microstructural configuration of the RVE \\
\hline
\multicolumn{2}{|>{\columncolor{gray!15}}c|}{\textbf{Operator Learning}} \\ \hline
$\mathcal{X}$ & The input function space \\ 
$\mathcal{Y}$ & The output function space \\ 
$u \in \mathcal{X}$ & Input function 
\\
$v \in \mathcal{Y}$ & Output function  \\
$y$ & Query coordinate in the input domain of $v$ \\
$\Phi: \mathcal{X} \rightarrow \mathcal{Y}$ & The operator mapping between function spaces 
\\
$\curlyE: \curlyX \to \R^n$ & Encoder mapping \\
$\curlyD: \R^n \to \curlyY $ & Decoder mapping \\
% $\mathcal{F}, \mathcal{F}^{-1}$ & Fourier transform and its inverse \\    
% $k$ & Fourier modes / wave numbers \\ 
% $k_{\text{max}}$ & Max Fourier modes used in the Fourier layer \\ 
$\phi$ & Activation function    \\
\hline
\multicolumn{2}{|>{\columncolor{gray!15}}c|}{\textbf{Micrometer}} \\ \hline
$\text{PE}$ &  Positional embedding \\ 
% $\text{SA}$ & Self-attention \\
$\text{MSA}$ & Multi-head self-attention \\
$\text{MHA}$ & Multi-head attention \\
% $\text{CSA}$ & Multi-head cross-attention \\
$\text{LN}$ & Layer normalization \\ 
$P$ & Patch size of Vision Transformer \\
$D$ & Embedding dimension of Vision Transformer \\
% $N_x \times N_y$ & Resolution of dummy grid  \\
% $\mathbf{x} \in \mathbb{R}^{N_x \times N_y \times C}$ & Latent grid features \\
$\beta$ & Locality of interpolated latent grid features  \\
\hline
\multicolumn{2}{|>{\columncolor{gray!15}}c|}{\textbf{Hyperparameters}} \\ \hline
$B$ & Batch size \\ 
$Q$ & Number of query coordinates in each batch \\ 
$H \times W$ & Resolution of spatial discretization \\
 \hline
\end{tabular}
\label{tab:notation}
\end{table}

\section{Green's function for Lippmann-Schwinger equation}
\label{math:LSequation}

\subsection{Periodical isotropic Green’s function in Fourier space}

Recall that Lippmann-Schwinger equation is given by
\begin{equation}
\varepsilon_{i j}(\mathbf{x})+\int_{\Omega} G_{i j k l}^{(0)}\left(\mathbf{x}, \mathbf{x}^{\prime}\right): \tau_{k l}\left(\mathbf{x}^{\prime}\right) \mathrm{d} \mathbf{x}^{\prime}-\bar{\varepsilon}_{i j}=0.
\end{equation}

To solve the above equations, one can choose strain or stress controlled PBC applied to the RVE
\begin{align}
    \frac{1}{|\Omega|} \int_{\Omega} \varepsilon_{ij} \mathrm{d} \mathbf{x} = \bar{\varepsilon}_{i j} = \varepsilon_{ij,M} \quad \text{or} \quad  \frac{1}{|\Omega|} \int_{\Omega} \sigma_{ij} \mathrm{d} \mathbf{x} = \sigma_{ij,M},
\end{align}
where $\varepsilon_{ij,M}$ and $\sigma_{ij,M}$ denote macroscopic homogenized strain and stress, respectively. Mixed boundary conditions can also be imposed in general. 

Here the Green's function $G_{i j k l}^{(0)}\left(\mathbf{x}, \mathbf{x}^{\prime}\right)$ is a non-local function, which represents the strain field at point $\mathbf{x}$ generated by a concentrated external stress field at point $\mathbf{x}^{\prime}$ in homogeneous reference material. Due to the periodic assumption of the RVE, $\mathbb{G}^{(0)}(\mathbf{x})$ can be expressed in Fourier space as  $\hat{\mathbb{G}}^{(0)}(\bm{\xi})$, which can be determined by either continuous or discrete scheme. $\bm{\xi}$ is the frequency vector in Fourier space corresponding to $\mathbf{x}$ in Euclidean space while the domain $\Omega$ is discretized with pixel grid, for instance:
\begin{equation}
\xi_i=\frac{2 \pi}{h_i T_i}
\left\{
\begin{aligned}
&\left[0,1, \ldots, \frac{T_i}{2}-1,-\frac{T_i}{2}, \ldots,-1\right], & \text { if } T_i \text { is even.}\\
&\left[0,1, \ldots, \frac{T_i-1}{2},-\frac{T_i-1}{2}, \ldots,-1\right], & \text { if } T_i \text { is odd.}
\end{aligned}
\right.
\label{setup:spa_discretized}
\end{equation}
where $i=1,2$ represents the indices of spatial coordinate in 2D. $h_i$ and $T_i$ denotes the pixel size and the number of pixels along each spatial coordinate, respectively.

In general, the continuous formulation of above isotropic Green's function $\hat{\mathbb{G}}^{(0)}(\bm{\xi})$ in Fourier space is written as:
\begin{equation}
\left\{
\begin{aligned}
\hat{G}_{i j k l}^{(0)}(\bm{\xi})&=0, \text { if } \bm{\xi} = \mathbf{0}, \\
\hat{G}_{i j k l}^{(0)}(\bm{\xi})=\frac{1}{4 \mu^{0}} \hat{G}_{i j k l}^{(1)}(\bm{\xi})+&\frac{\mu^{0}+\lambda^{0}}{\mu^{0}\left(2 \mu^{0}+\lambda^{0}\right)} \hat{G}_{i j k l}^{(2)}(\bm{\xi}), \text { if } \bm{\xi} \neq \mathbf{0}.
\end{aligned}
\right.
\label{setup:Goperator}
\end{equation}
\noindent
$\hat{G}_{i j k l}^{(1)}(\bm{\xi})$ and $\hat{G}_{i j k l}^{(2)}(\bm{\xi})$ can be expressed as follows:
\begin{equation}
\hat{G}_{i j k l}^{(1)}(\bm{\xi})=\frac{\delta_{i k} \xi_{j} \xi_{l}+\delta_{i l} \xi_{j} \xi_{k}+\delta_{j k} \xi_{i} \xi_{l}+\delta_{j l} \xi_{i} \xi_{k}}{\|\bm{\xi}\|^{2}}, \quad  \hat{G}_{i j k l}^{(2)}(\bm{\xi})=-\frac{\xi_{i} \xi_{j} \xi_{k} \xi_{l}}{\|\bm{\xi}\|^{4}}.
\label{setup:Goperator12}
\end{equation}
The mathematical proof of Eq.\eqref{setup:Goperator} and Eq.\eqref{setup:Goperator12}  is provided in Appendix \ref{appendix:proof}.
% $\lambda^{0}$ and $\mu^{0}$ are Lame constants of elastic homogeneous reference material. The selection of $\lambda^{0}$ and $\mu^{0}$ will be discussed in Section \ref{sec:data_generation}.

\subsection{Mathematical proof}
\label{appendix:proof}
The aim of this subsection is to find the Green's function $\mathbb{G}^0$ corresponding to isotropic homogeneous reference material $\mathbb{C}^0$ in the Lippmann-Schwinger equation (see Eq.(\ref{setup:dotLS})). Recalling that the periodic boundary condition (PBC) is applied to the RVE, the periodical physical fields can be denoted by Neumann series with a single wave vector \cite{Mura1982book,Li2008book}, such as:
\begin{equation}
\square^{*}(\mathbf{x})=\hat{\square}^{*}(\boldsymbol{\xi}) \exp (\mathrm{i} \boldsymbol{\xi} \cdot \mathbf{x}),
\label{singlewave}
\end{equation}
where $\mathrm{i}$ denotes to $\sqrt{-1}$ and $\square^{*}$ can be referred to $\boldsymbol{\sigma}^{*}, \boldsymbol{\varepsilon}^{*}, \boldsymbol{\tau}^{*}$ and $\boldsymbol{s}^{*}$. Substitute Eq.(\ref{setup:polarstress}) into Eq.(\ref{setup:balanceequation}) and express each physical field periodically, we can arrive at:
\begin{equation}
\begin{dcases*}
\frac{\partial \sigma^{*}_{i j}(\mathbf{x})}{\partial x_i}=0, \\
\sigma^{*}_{i j}(\mathbf{x})=C^0_{i j k l}(\mathbf{x}): \varepsilon^{*}_{k l}(\mathbf{x}) + \tau_{ij}^{*}(\mathbf{x}),\\
\varepsilon^{*}_{i j}(\mathbf{x})=\frac{1}{2}\left(\frac{\partial s^{*}_{i}(\mathbf{x})}{\partial 
 x_j}+\frac{\partial s^{*}_{j}(\mathbf{x})}{\partial 
 x_i}\right).
\end{dcases*}
\label{Periodicbalanceequation}
\end{equation}
Recalling that the strain field and displacement field can be decomposed into an average and fluctuation part, we have
\begin{equation}
\begin{split}
\left\{\begin{array}{lc}
s^{*}_{i}(\mathbf{x})=\tilde{s}^{*}_{i}(\mathbf{x})+\bar{s}_{i}, \\
\varepsilon^{*}_{i j}(\mathbf{x})=\tilde{\varepsilon}^{*}_{i j}(\mathbf{x})+\bar{\varepsilon}_{i j}.
\end{array}
\right.
\end{split}
\end{equation}

By Substituting Eq.(\ref{singlewave}) into Eq.(\ref{Periodicbalanceequation}) and eliminating $\sigma^{*}_{i j}$, we have
\begin{equation}
C_{ijkl}^{0} \xi_{j} \xi_{l} \hat{\tilde{s}}_{k}^{*}(\boldsymbol{\xi})=\mathrm{i} \hat{\tau}^{*}_{ij}(\boldsymbol{\xi}) \xi_{j}.
\label{Fourierbalance}
\end{equation}
Denote the coefficient of $\hat{\tilde{s}}_{k}^{*}$ in the left hand side of above equation by:
\begin{equation}
    K_{i k}(\boldsymbol{\xi})=C_{i j k l}^{0} \xi_{l} \xi_{j},
\end{equation}
which is the acoustic tensor of the homogeneous reference material. Since the fourth order elastic tensor $C_{ijkl}^{0}$ can be expressed as:
\begin{equation}
C_{i j k l}^{0}=\lambda^{0} \delta_{i j} \delta_{k l}+\mu^{0}\left(\delta_{i k} \delta_{j l}+\delta_{i l} \delta_{j k}\right).
\end{equation}
Therefore, $K_{i k}(\boldsymbol{\xi})$ can be expressed as:
\begin{equation}
    K_{i k}(\boldsymbol{\xi}) = (\mu^{0}+\lambda^{0})\xi_{i} \xi_{k} + \mu^{0}\|\boldsymbol{\xi}\|^{2}\delta_{i k}.
\end{equation}
We can denote the inverse of $K_{i k}(\boldsymbol{\xi})$ by $N_{ki}(\boldsymbol{\xi})$, such as:
\begin{equation}
N_{ki}(\boldsymbol{\xi})=\frac{1}{\mu^{0}\|\boldsymbol{\xi}\|^{2}}\left(\delta_{ki}-\frac{\mu^{0}+\lambda^{0}}{\mu^{0}+2 \lambda^{0}} \frac{\xi_{k} \xi_{i}}{\|\boldsymbol{\xi}\|^{2}}\right).
\label{inverseacustic}
\end{equation}
Using the above equation, $\hat{\tilde{s}}_{k}^{*}$ can be expressed as:
\begin{equation}
    \hat{\tilde{s}}_{k}^{*}(\boldsymbol{\xi})=\mathrm{i} N_{k i}(\boldsymbol{\xi})  \hat{\tau}^{*}_{i j}(\boldsymbol{\xi})\xi_{j}.
\end{equation}
According to the symmetrical property of polarization stress $\hat{\tau}^{*}_{i j}$, $\hat{\tilde{s}}_{k}^{*}$ can be further denoted by:
\begin{equation}
\hat{\tilde{s}}_{k}^{*}(\boldsymbol{\xi})=\frac{\mathrm{i}}{2}\left(N_{k i}(\boldsymbol{\xi}) \xi_{j}+N_{k j}(\boldsymbol{\xi}) \xi_{i}\right) \hat{\tau}^{*}_{i j}(\boldsymbol{\xi}).
\end{equation}
Taking the symmetric gradient of $\hat{\tilde{s}}_{k}^{*}$, the strain fluctuation can be expressed as:
\begin{equation}
\hat{\tilde{\varepsilon}}_{kl}^{*}(\boldsymbol{\xi})=-\frac{1}{4}\left(N_{k i}(\boldsymbol{\xi}) \xi_{j} \xi_{l}+N_{k j}(\boldsymbol{\xi}) \xi_{i} \xi_{l}+N_{li}(\boldsymbol{\xi}) \xi_{j} \xi_{k}+N_{lj}(\boldsymbol{\xi}) \xi_{i} \xi_{k}\right) \hat{\tau}^{*}_{i j}(\boldsymbol{\xi}).
\end{equation}
The Green's function is defined as a symmetric fourth order tensor which creates a mapping from polarization stress $\hat{\tau}^{*}_{i j}(\boldsymbol{\xi})$ to strain fluctuation $\hat{\tilde{\varepsilon}}_{kl}^{*}(\boldsymbol{\xi})$:
\begin{equation}
\hat{G}_{klij}^{0}(\boldsymbol{\xi})=-\frac{1}{4}\left(N_{k i}(\boldsymbol{\xi}) \xi_{j} \xi_{l}+N_{k j}(\boldsymbol{\xi}) \xi_{i} \xi_{l}+N_{li}(\boldsymbol{\xi}) \xi_{j} \xi_{k}+N_{lj}(\boldsymbol{\xi}) \xi_{i} \xi_{k}\right).
\end{equation}
Substitute Eq.(\ref{inverseacustic}) into above equation, the Green's function can be computed by:
\begin{equation}
\hat{G}_{i j k l}^{0}(\boldsymbol{\xi})=\frac{1}{4 \mu^{0}\|\boldsymbol{\xi}\|^{2}}\left(\delta_{k i} \xi_{j} \xi_{l}+\delta_{i l} \xi_{j} \xi_{k}+\delta_{j l} \xi_{i} \xi_{k}+\delta_{j k} \xi_{i} \xi_{l}\right)-\frac{\left(\mu^{0}+\lambda^{0}\right)}{\mu^{0}\left(2 \mu^{0}+\lambda^{0}\right)} \frac{\xi_{i} \xi_{j} \xi_{k} \xi_{l}}{\|\boldsymbol{\xi}\|^{4}},
\end{equation}
which is corresponding to Eq.(\ref{setup:Goperator}) and Eq.(\ref{setup:Goperator12}). Therefore, the proof ends.

\subsection{Expression of Green's function in Fourier space by Voigt notation}
Following Eq.(\ref{setup:Goperator12}) in Section \ref{math_theory}, the Green's function in Fourier Space is expressed by:
\begin{equation}
        \hat{\mathbb{G}}^{(0)}(\boldsymbol{\xi})=\frac{1}{4 \mu^{0}} \hat{\mathbb{G}}^{(1)}(\boldsymbol{\xi})+\frac{\left(\mu^{0}+\lambda^{0}\right)}{\mu^{0}\left(2 \mu^{0}+\lambda^{0}\right)} \hat{\mathbb{G}}^{(2)}(\boldsymbol{\xi}).
\end{equation}
Each component in $\hat{\mathbb{G}}^{(1)}(\boldsymbol{\xi})$ and $\hat{\mathbb{G}}^{(2)}(\boldsymbol{\xi})$ in terms of Voigt notation is given by:
\begin{equation}
\hat{\mathbb{G}}^{(1)}(\boldsymbol{\xi})=\left[\begin{array}{ccc}\frac{4 \xi_{1}{ }^{2}}{\|\boldsymbol{\xi}\|^{2}} & 0 & \frac{4 \xi_{1} \xi_{2}}{\|\boldsymbol{\xi}\|^{2}}\\0 & \frac{4 \xi_{2}{ }^{2}}{\|\boldsymbol{\xi}\|^{2}} & \frac{4 \xi_{1} \xi_{2}}{\|\boldsymbol{\xi}\|^{2}} \\ \frac{4 \xi_{1} \xi_{2}}{\|\boldsymbol{\xi}\|^{2}} & \frac{4 \xi_{1} \xi_{2}}{\|\boldsymbol{\xi}\|^{2}}  & \frac{4\left(\xi_{1}{ }^{2}+\xi_{2}{ }^{2}\right)}{\|\boldsymbol{\xi}\|^{2}} \end{array}\right],
\end{equation}

\begin{equation}
\hat{\mathbb{G}}^{(2)}(\boldsymbol{\xi})=-\left[\begin{array}{ccc}
\frac{\xi_{1}{ }^{4}}{\|\boldsymbol{\xi}\|^{4}} & \frac{\xi_{1}{ }^{2} \xi_{2}{ }^{2}}{\|\boldsymbol{\xi}\|^{4}}  & \frac{2 \xi_{1}{ }^{3} \xi_{2}}{\|\boldsymbol{\xi}\|^{4}} \\
\frac{\xi_{1}{ }^{2} \xi_{2}{ }^{2}}{\|\boldsymbol{\xi}\|^{4}} & \frac{\xi_{2}{ }^{4}}{\|\boldsymbol{\xi}\|^{4}} & \frac{2 \xi_{1} \xi_{2}{ }^{3}}{\|\boldsymbol{\xi}\|^{4}}  \\
\frac{2 \xi_{1}{ }^{3} \xi_{2}}{\|\boldsymbol{\xi}\|^{4}} & \frac{2 \xi_{1} \xi_{2}{ }^{3}}{\|\boldsymbol{\xi}\|^{4}} &  \frac{4 \xi_{1}{ }^{2} \xi_{2}{ }^{2}}{\|\boldsymbol{\xi}\|^{4}} 
\end{array}\right].
\end{equation}

\section{Data Generation}

\subsection{FFT based homogenization.}
\label{appendix: fft}

Here we summarize the numerical method and process for generating the outputs of  training and test dataset, typically, the point-wise strain concentration tensor $\mathbb{A}(\mathbf{x})$ for each RVE. Throughout this work, these datasets are generated via solving Eq.\eqref{setup:convolutionLS} by fast Fourier transformation (FFT) based homogenization \cite{Moulinec1998FFT}, which is an accurate and efficient numerical method. The systematic solution scheme is provided in Algorithm \ref{setup:FFTsolver}. Compared to other element based numerical methods, the convolution term between Green's function and polarization stress is transferred into dot product and computed in Fourier space via discrete Fourier transform (DFT) rather than in Euclidean space, which exhibits more time and memory efficiency. Rather, the point-wise stress is updated in Euclidean space by local constitutive law. Concerning the convergence test in Algorithm\ref{setup:FFTsolver}, we check the equilibrium state of $\boldsymbol{\sigma}^{(n)}(\mathbf{x})$, referred to the local stress field $\boldsymbol{\sigma}(\mathbf{x})$ at iteration counter $n$. Following the same principle in references \cite{Moulinec1998FFT,Willot2015FD}, an error index $\text{Tol}^n$ to check convergence is defined as:
\begin{equation}
    \text{Tol}^{n}=\sqrt{\frac{\sum_{\bm{\xi}}\left|\bm{\xi} \cdot \hat{\boldsymbol{\sigma}}^{(n)}(\bm{\xi})\right|^{2}}{T_1T_2[\hat{\sigma}_{ij}^{(n)}(\mathbf{0}):\hat{\sigma}_{ij}^{(n)}(\mathbf{0})]}}{\ll}1.
    \label{FFTcvtest}
\end{equation}
By virtue of Parseval's theorem \cite{stein2009fourier}, the convergence criteria can be checked in either Euclidean space or Fourier space. $\hat{\boldsymbol{\sigma}}^{(n)}(\mathbf{0})$ refers to stress in Fourier space at zero frequency, corresponding to the macroscopic average stress in Euclidean space. The fixed-point iteration stops while $\text{Tol}^{n}$ reaches a prescribed value (e.g., $10^{-6}$). The convergence rate of the fixed-point iteration solver in Algorithm \ref{setup:FFTsolver} is highly dependent on the choice of reference material $\mathbb{C}^{{0}}$, which is determined by $\lambda_0$ and $\mu_0$. Following the work of Moulinec and Suquet \cite{Moulinec1998FFT}, to achieve the best convergence property, a pair of $\lambda_0$ and $\mu_0$ is chosen as
\begin{equation}
\left\{
\begin{aligned}
\lambda^0=\frac{1}{2}\left(\underset{\mathbf{x} \in \Omega}{\text{inf}} \lambda(\mathbf{x})+\underset{\mathbf{x} \in \Omega}{\text{sup}}  \lambda(\mathbf{x})\right), \\
\mu^0=\frac{1}{2}\left(\underset{\mathbf{x} \in \Omega}{\text{inf}}\mu(\mathbf{x})+\underset{\mathbf{x} \in \Omega}{\text{sup}}  \mu(\mathbf{x})\right),
\end{aligned}
\right.
\label{setup:C0select}
\end{equation}
where $\lambda(\mathbf{x})$ and $\mu(\mathbf{x})$ are computed through $E(\mathbf{x})$ and $\nu(\mathbf{x})$ following Eq.\eqref{setup:Lametransform}.
Besides, the usage of a continuous Green's function in FFT based homogenization yields numerical defects in local physical fields, such as Ringing artifacts, stress oscillations, to name few \cite{TongruiIJSS,Schneider2020review}. To alleviate those negative effects, here we adopt the discrete Green's function based on modified frequency vectors, which can be derived by using the finite difference method (FDM) on a rotated grid. The modified frequency vectors $\tilde{\boldsymbol{\xi}}$ are expressed as \cite{Gelebart2020FFT} 
\begin{equation}
\left\{\begin{aligned}
\tilde{\xi}_{1}=\frac{2}{h_{1}} \sin \left(\frac{\xi_{1}}{2}\right) \cos \left(\frac{\xi_{2}}{2}\right), \\
\tilde{\xi}_{2}=\frac{2}{h_{2}} \cos \left(\frac{\xi_{1}}{2}\right) \sin \left(\frac{\xi_{2}}{2}\right),\\
\end{aligned}\right.
\end{equation}
where $h_i (i = 1,2)$ denotes the pixel size along direction $i$. The numerical implementation is achieved by replacing $\boldsymbol{\xi}$ with $\tilde{\boldsymbol{\xi}}$ in Eq.\eqref{setup:Goperator}. For a mathematical proof of the discrete Green's function, interested readers are referred to  \cite{Willot2015FD,Gelebart2020FFT,Schneider2020review} for details.

% \paragraph{Dataset generation pipeline.}
% This section presents the algorithm of the dataset generation process, including the RVE generation, material proprieties sampling and calculation of strain concentration tensor. The algorithm is given in Algorithm \ref{algo:datageneration} in pseudo code format.

\begin{algorithm}[h]
\setstretch{1.4}
\SetAlgoLined
\caption{Basic scheme of FFT based homogenization \cite{Moulinec1998FFT}}
$\mathbf{Step.1}$: Initialization, set the strain controlled PBC using Eq.\eqref{setup:macroBC}. For each $\mathbf{x} \in \Omega$, initial local strain and stress fields are set as:
$$
\boldsymbol{\varepsilon}^{(0)}(\mathbf{x})=\overline{\boldsymbol{\varepsilon}} =\boldsymbol{\varepsilon}_{M}, \quad \boldsymbol{\sigma}^{(0)}(\mathbf{x})= \mathbb{C}(\mathbf{x}): \boldsymbol{\varepsilon}^{(0)}(\mathbf{x}).
$$

$\mathbf{Step.2}$: Solve Eq.\eqref{setup:convolutionLS} by fixed point iteration to obtain $\boldsymbol{\varepsilon}^{(n+1)}(\mathbf{x}), \boldsymbol{\sigma}^{(n+1)}(\mathbf{x})$.

\hspace{0.3cm} a. Compute the polarization stress in Euclidean space: $\boldsymbol{\tau}^{(n)}(\mathbf{x})=\boldsymbol{\sigma}^{(n)}(\mathbf{x})-\mathbb{C}^{{0}}: \boldsymbol{\varepsilon}^{(n)}(\mathbf{x})$.

\hspace{0.3cm} b. Fast Fourier transform on polarization stress: $\hat{\boldsymbol{\tau}}^{(n)}(\boldsymbol{\xi})=\mathcal{F}\left(\boldsymbol{\tau}^{(n)}(\mathbf{x})\right)$.

\hspace{0.3cm} c. Compute local strain in Euclidean space: $\boldsymbol{\varepsilon}^{(n+1)}(\mathbf{x})=\overline{\boldsymbol{\varepsilon}}-\mathcal{F}^{-\mathbf{1}}\left(\hat{\mathbb{G}}^{({0})}(\boldsymbol{\xi}): \hat{\boldsymbol{\tau}}^{(n)}(\boldsymbol{\xi})\right)$.

\hspace{0.3cm} d. Update local stress in Euclidean space: $\boldsymbol{\sigma}^{(n+1)}(\mathbf{x})=\mathbb{C}(\mathbf{x}): \boldsymbol{\varepsilon}^{(n+1)}(\mathbf{x})$.

$\mathbf{Step.3}$: Convergence test using Eq.\eqref{FFTcvtest}, if not convergence, repeating $\mathbf{Step.2}$ with iteration ${n} = {n}+{1}$.
\label{setup:FFTsolver}
\end{algorithm}

\begin{algorithm}[h]
\setstretch{1.4}
\SetAlgoLined
\caption{Pipeline of data generation}
% Set the sampling size (number of RVE domain) as $N_p$. The iteration counter below is set to be $i$, with $1\leq i \leq N_p$. 
\KwIn{$N_p$: Number of RVE domains, $T_1 \times T_2$: RVE resolutions, $R_d$: Fiber radius, $Std$: Standard deviation of fiber radius}
\KwOut{$\textit{RVE\_image}$, $\textit{labels\_mat}$, $A_{ten}$}
\BlankLine
$\mathbf{Step.1}$: \textbf{Sampling of material 
properties}
\BlankLine
\Begin{
{a. Generate material properties for each RVE and store them in an array $\textit{labels\_mat} \in \mathbb{R}^{N_p \times 4}$ using Latin Hypercube Sampling (LHS), where row $i$ contains $[E_f^i, \nu_f^i, E_m^i, \nu_m^i]$ for $\Omega^i$:

[$\textit{labels\_mat}$] = \texttt{LHS\_design}($N_p$, $E_{f,max}$, $\nu_{f,max}$, $E_{m,max}$, $\nu_{m,max}$, $E_{f,min}$, $\nu_{f,min}$, $E_{m,min}$, $\nu_{m,min}$)

{where $E_{\square,\square}$ and $\nu_{\square,\square}$ are referred to the upper and lower bounds of Young's modulus and Poisson ratio for  matrix and fiber materials, respectively. These ranges of material properties are summarized in Table\ref{tab:dataset}}.
}
}
\BlankLine
$\mathbf{Step.2}$: \textbf{RVE and fourth order elastic tensor generation}
\BlankLine
\Begin{
{a. Initialize a 1D array $\textit{labels\_Vof}$ to store the fiber volume fraction $\textit{Vof}^i$ for all $\Omega^i$.

b. Divide $N_p$ into 20 groups with fiber volume fraction uniformly distributed ranging from 40\% to 60\%. 

c. Set array $\textit{labels\_Vof}$:} \quad [$\textit{labels\_Vof}$] = \texttt{reshape} $(\texttt{repmat}(40:1:60, N_p/20, 1), N_p, 1)$

{d. Initialize a 3D array $\textit{RVE\_image} \in \R^{T_1 \times T_2 \times N_p}$ to store the RVE information, where $\textit{RVE\_image}(:,:,i)$ is a boolean array representing the $i$-th RVE's microstructure (0: matrix, 1: fiber). Meanwhile, initialize a 5D array $C_{ten} \in \R^{T_1 \times T_2 \times 3 \times 3 \times N_p}$ to store the fourth order elastic tensor, where $C_{ten}(:,:,:,:,i)$  corresponds to the pixel-wise fourth order elastic tensor for $i$-$\text{th}$ RVE.

e. Generate RVEs and fourth order elastic tensor, and store them in $\textit{RVE\_image}$ and $C_{ten}$, respectively:}

\For{$i \in [1, N_p]$}{
    $\textit{RVE\_image}(:,:,i)$ = $\texttt{RAND{\_}uSTRU{\_}GEN}$ ($R_d$, $\textit{labels\_Vof}(i)$, $Std$, $T_1$, $T_2$)

       \For{$p \in [1, T_1]$}{
\For{$q \in [1, T_2]$}{
   $\left[C_{ten}(p,q,:,:,i)\right]$ = \texttt{C{\_}tensor}($\textit{RVE\_image}(p,q,i)$, $\textit{labels\_mat}(i,:)$) \texttt{\# Eqs.\eqref{setup:Lametransform} and \eqref{setup:Lameconstant}}
   }
   }
  }
}
\BlankLine
$\mathbf{Step.3}$: \textbf{Calculation of strain concentration tensor}
\BlankLine
\Begin{
{a. Initialize a 5D array $A_{ten} \in \R^{T_1 \times T_2 \times 3 \times 3 \times N_p}$ to store the strain concentration tensor, where $A_{ten}(:,:,:,:,i)$  corresponds to the full-field strain concentration tensor for $i$-$\text{th}$ RVE. 

b. Run Algorithm \ref{setup:FFTsolver} to compute $A_{ten}$:}

\For{$i \in [1, N_p]$}{
   $\left[A_{ten}(:,:,:,:,i)\right]$ = \texttt{FFT{\_}homogenization}$\left(C_{\textit{ten}}(:,:,:,:,i)\right)$
}
}
\label{algo:datageneration}
\end{algorithm}

\clearpage
\section{FE-Micrometer}

Algorithm \ref{algo:FE_micrometer} outlines the framework of FE-Micrometer for concurrent multiscale modeling.

\begin{algorithm}[h]
\setstretch{1.2}
\small
\SetAlgoLined
\caption{Concurrent multiscale modeling using FE-Micrometer}
\BlankLine
$\mathbf{Step.1}$: \textbf{Prepare the macroscale FE mesh, material properties and model inputs \& outputs}
\BlankLine
\Begin{
{{a. Generate the marcoscale FE mesh with $N_{node}$ nodes and set $N_p$ as the number of RVEs in FE mesh.}

{b. Set macroscale material properties (for each RVE domain $\Omega^i$)  by using Eq.\eqref{klexpansion}, and store them in an array $\textit{labels\_mat} \in \mathbb{R}^{N_p \times 4}$, where row $i$ contains $[E_f^i, \nu_f^i, E_m^i, \nu_m^i]$ for $\Omega^i$.}

{c. Following Step.2 in Algorithm \ref{algo:datageneration}, generate RVEs and fourth order elastic tensor, and store them in $\textit{RVE\_image} \in \R^{T_1 \times T_2 \times N_p}$ and $C_{ten} \in \R^{T_1 \times T_2 \times 3 \times 3 \times N_p}$, respectively.}

{d. Inference the \textit{Micrometer} and store the predicted strain concentration tensor into $A_{ten} \in \R^{T_1 \times T_2 \times 3 \times 3 \times N_p}$:
% \begin{align}
%     \left[A_{ten}\right] = \texttt{Mircometer}\left(C_{\textit{ten}}\right)
% \end{align}
\begin{center}
    $\left[A_{ten}\right] = \texttt{Mircometer}\left(C_{\textit{ten}}\right)$
\end{center}
}
}
}
\BlankLine
$\mathbf{Step.2}$: \textbf{Quasi static analysis: FE-Micrometer two-scale calculation at arbitrary time interval} $\boldsymbol{\left[t_{n},t_{n+1}\right]}$

\KwIn{$\boldsymbol{s}^{n} \in \R^{2N_{node}}$ and $\mathbb{C}_{M,tan}^{n} \in \R^{3\times 3 \times N_p}$: Displacement and macroscale consistent tangent stiffness at $t_{n}$, $A_{ten}$: strain concentration tensor, $C_{ten}$: fourth order elastic tensor, $\boldsymbol{F}_{ext}^{n} \in \R^{2N_{node}}$: External force, $\textit{Dof\_l}$: label of DOFs with loading, $\textit{Dof\_f}$ : label of DOFs in stress free status}
\KwOut{$\boldsymbol{s}^{n} \in \R^{2N_{node}}$, $\mathbb{C}_{M,tan}^{n} \in \R^{3\times 3 \times N_p}$, $\boldsymbol{\sigma}_M^{n+1} \in \R^{3\times 1 \times N_p}$, $\boldsymbol{\varepsilon}_M^{n+1} \in \R^{3\times 1 \times N_p}$, $\boldsymbol{\sigma}^{n+1} \in \R^{T_1 \times T_2 \times 3 \times 1 \times N_p}$, $\boldsymbol{\varepsilon}^{n+1} \in \R^{T_1 \times T_2 \times 3 \times 1 \times N_p}$ and $\boldsymbol{F}_{int}^{n+1} \in \R^{2N_{node}} $: Displacement, macroscale consistent tangent stiffness, macrostress, macrostrain, microstress, microstrain and internal force at $t_{n+1}$}
\BlankLine
\Begin{
Set $\boldsymbol{s}^{n,k} \leftarrow \boldsymbol{s}^{n}$, $\mathbb{C}_{M,tan}^{n,k} \leftarrow \mathbb{C}_{M,tan}^{n}$ and apply the macroscale boundary conditions.

\Repeat{$|\boldsymbol{R}^{n,k}(\textit{Dof\_f})|_{2}\leq \texttt{TOL}$, \text{with} $\texttt{TOL} = 10^{-7}$ }
{{a. Compute the macroscale global stiffness matrix $\mathbf{K}_u$ and internal force vector $\boldsymbol{F}_{int}$ using macroscale consistent tangent stiffness $\mathbb{C}_{M,tan}^{n,k}$ and displacement field $\boldsymbol{s}^{n,k}$.}

{b. Get displacement field $\boldsymbol{s}^{n,k+1}$ and macrostrain $\boldsymbol{\varepsilon}_M^{n,k+1}$:}
\begin{equation}
    \boldsymbol{R}^{n,k+1} = \boldsymbol{F}_{ext}^{n}-\boldsymbol{F}_{int}, \quad
    \boldsymbol{s}^{n,k+1} = \boldsymbol{s}^{n,k}-\left[\mathbf{K}_u\right]^{-1}\boldsymbol{R}^{n,k+1} \quad \text{and} \quad \boldsymbol{\varepsilon}_M^{n,k+1} = \nabla^{\text{sym}} \boldsymbol{s}^{n,k+1}
\end{equation}\
{c. Get microstress $\boldsymbol{\sigma}^{n,k+1}$, microstrain $\boldsymbol{\varepsilon}^{n,k+1}$, macrostress $\boldsymbol{\sigma}_{M}^{n,k+1}$ and consistent tangent stiffness $\mathbb{C}_{M,tan}^{n,k+1}$. The function \texttt{Mat} below corresponds to Eqs. \eqref{setup:Lametransform}, \eqref{setup:Lameconstant} and \eqref{eq:Atensor}:

\For{$i \in [1, N_p]$}{ 

\For{$p \in [1, T_1]$}{
\For{$q \in [1, T_2]$}{
   $\left[\boldsymbol{\sigma}^{n,k+1}(p,q,:,:,i),\boldsymbol{\varepsilon}^{n,k+1}(p,q,:,:,i)\right]$ = \texttt{Mat}($C_{\textit{ten}}(p,q,:,:,i)$,$A_{\textit{ten}}(p,q,:,:,i)$,$\boldsymbol{\varepsilon}_{M}^{n,k+1}(:,:,i)$)
   }
   }
Compute the macrostress $\boldsymbol{\sigma}_{M}^{n,k+1}(:,:,i)$ by averaging microstress $\boldsymbol{\sigma}^{n,k+1}(:,:,:,:,i)$ in $\Omega_i$

Compute the macroscale tangent stiffness $\mathbb{C}_{M,tan}^{n,k+1}(:,:,i)$ by averaging the pagewise production between $C_{\textit{ten}}(p,q,:,:,i)$ and $A_{\textit{ten}}(p,q,:,:,i)$ in $\Omega_i$, see Eq.\eqref{homogenizationrule} 
}

Compute $\boldsymbol{F}_{int}^{n+1,k}$ and get the reaction force by summing up $\boldsymbol{F}_{int}^{n+1,k}$($\textit{Dof\_l}$)

Set  $k\leftarrow k+1$
}

}
}
Updating the solution: Set $\boldsymbol{s}^{n,k} \rightarrow \boldsymbol{s}^{n+1}$
\label{algo:FE_micrometer}
\end{algorithm}

\clearpage
\section{Overview of Fourier Neural Operator (FNO)}
\label{app: fno}

Integral Kernel Neural Operators form a general class of neural operators, which learn mappings between function spaces by approximating the integral kernel of an operator using neural networks. They were introduced as a generalization of the Neural Operator framework proposed by Li {\it et al.} \cite{li2020neural}.
The key idea here is to represent the underlying operator $\Phi: \mathcal{X} \rightarrow \mathcal{Y}$ as a convolutional integral operator, $\Phi(u)(y) = \int_{\Omega} \kappa(x - y) u(x) dx$,
where $\kappa$ is a kernel function.
In practice, the kernel function $\kappa$ is parameterized by a neural network, and the integral is then approximated using a quadrature rule, such as Monte Carlo integration, Gaussian quadrature, or via spectral convolution in the frequency domain.

FNO is one representative example of this paradigm, defined as:
\begin{align}
\Phi(u ; \theta) & =\mathcal{Q} \circ \mathcal{H}_L \circ \cdots \mathcal{H}_2 \circ \mathcal{H}_1 \circ \mathcal{R}(u), 
\end{align}
where each hidden layer $\mathcal{H}_{\ell}(v)(x ; \theta_\ell)$ is given by:
\begin{align}
\mathcal{H}_{\ell}(v)(x ; \theta) & = {\phi} \left(W_{\ell} v(x)+b_{\ell}+\mathcal{K}(v)\left(x ; \theta_{\ell}\right)\right),
\end{align}
with a component-wise activation function $\phi$. 
The architecture consists of input and output layers, $\mathcal{R}$ and $\mathcal{Q}$, respectively, which are typically pointwise compositions using either shallow neural networks or linear transformations. 
The hidden layers  $\mathcal{H}_\ell$  involves an affine
transformation and incorporates a convolutional integral operator $\mathcal{K}$, parameterized by $\theta_\ell$. This operator can be efficiently evaluated using the Fourier transform $\mathcal{F}$, resulting in a matrix-valued Fourier multiplier:
\begin{align}
    \mathcal{K}(v)\left(x ; \theta_{\ell}\right)=\int \kappa\left(x-y ; \theta_{\ell}\right) v(y) d y = \mathcal{F}^{-1}\left(\mathcal{F}\left(\kappa\left(\cdot ; \theta_{\ell}\right)\right) \mathcal{F}(v)\right),
\end{align}
where the Fourier transform $\mathcal{F}$ is computed component-wise.
More specifically, if $\kappa(x)=\left(\kappa_{i j}(x)\right)_{i j=1}^{d_c}$ represents the components of $\kappa(x)$, and $\widehat{\kappa}_{k,ij}$ denotes the $k$-th Fourier coefficient of $\kappa_{ij}(x)$, the $i$-th component $\mathcal{K}(v)_i$ of the output function $\mathcal{K}(v)$ is expressed as:
\begin{align}
    \left[\mathcal{K}(v)_i\right]\left(x ; \theta_{\ell}\right)=\frac{1}{(2 \pi)^d} \sum_{k \in \mathbb{Z}^d}\left(\sum_{j=1}^{d_c} \widehat{\kappa}_{k, i j} \mathcal{F}\left(v_j\right)(k)\right) e^{\mathrm{i} k \cdot x}.
\end{align}
Here the Fourier coefficients $\widehat{\kappa}_{k,ij}$ are the trainable parameters of the convolutional operator. In practice, a Fourier cut-off $k_{\max}$ is introduced, restricting a finite sum over $k$ up to wavenumbers $|k|_{\ell^\infty} \leq k_{\max}$, where $|\cdot|_{\ell^\infty}$ is the $\ell^{\infty}$-norm. This results in a finite set of trainable parameters, $\theta_{\ell} = \left\{\widehat{\kappa}_{k,ij} : |k|_{\ell^\infty} \leq k_{\max}, i,j=1,\dots,d_c\right\}$.

While the FNO is theoretically formulated directly on the function space without reduction to a finite-dimensional latent space, practical implementations typically discretize it by identifying the input and output functions with their point-values on an equidistant grid. This allows for efficient evaluation of the discrete Fourier transform using the fast Fourier transform algorithm (FFT), facilitating straightforward implementation in popular deep learning libraries.

\clearpage
\section{Model Details}
\label{appendix: baselines}

% \paragraph{Micrometer}

% We will use a 4 layers FNO encoders and 2 layers decoder for all configurations

\paragraph{Fourier Neural Operator (FNO)}  We employ a Fourier Neural Operator (FNO) model with four layers, where the number of channels increases from 32 to 128, and the Fourier modes range from 16 to 32. The GeLU activation function is employed, and no normalization scheme is applied. As shown in Table \ref{tab:different_models}, the  performance of FNO is inferior to that of Micrometer, even when using a comparable number of parameters.

\paragraph{Vision Transformer (ViT)} We employ various configurations of the Vision Transformer (ViT), as detailed in Table \ref{tab:vit_models}, and explore different patch sizes of 8 and 16. As shown in Table \ref{tab:different_models}, the model  performance is highly sensitive to the patch size, with smaller patches resulting in significantly better test errors. However, even with these adjustments, the accuracy of ViT remains lower than that of Micrometer when comparing models with similar numbers of parameters.

\begin{table}[h]
    % \small
      \renewcommand{\arraystretch}{1.2}
    \centering
    \caption{Details of Vision Transformer model variants.}
    \begin{tabular}{lcccccc}
        \toprule
        \textbf{Model} &   \textbf{Encoder layers}  &  \textbf{Embedding dim} & \textbf{MLP width} & \textbf{Heads} & \textbf{\# Params} \\
        \midrule
        ViT-S &  6  & 384 & 384 & 6 & 8 M  \\
        ViT-B &  12  & 512 & 1024 & 16 & 26 M  \\
        ViT-L &  18  & 768 & 1536 & 12 & 87 M  \\
        \bottomrule
    \end{tabular}
    \label{tab:vit_models}
\end{table}

\paragraph{UNet:} We adhere to the conventional implementation of the conventional UNet \cite{ronneberger2015u}, with group normalization. We vary the  embedding dimension of UNets, ranging from 32 to 128. As shown in Table \ref{tab:different_models}, the UNet consistently underperforms compared to the other models.

\clearpage
\section{Additional Results}

\begin{table}[h]
\renewcommand{\arraystretch}{1.3}
    \centering
        \caption{Performance of various deep learning models for predicting full microstress fields. Models include U-Net, Fourier Neural Operator (FNO), Vision Transformer (ViT), and Micrometer with different configurations.  Metrics include relative $L^1$ and $L^2$ errors, Root Mean Square Error (RMSE). Lower values ($\downarrow$) indicate better performance for error metrics.}
  \begin{tabular}{l c c c c c c}
        \toprule
        \textbf{Model} & \textbf{\# Params} & \textbf{Rel. $L^1$ error ($\downarrow$)}  & \textbf{Rel. $L^2$ error ($\downarrow$)}   & \textbf{RMSE ($\downarrow$)} & \textbf{Training time (hours)}\\
        \midrule
        UNet-16 & 2 M & 16.33\% & 16.99\% & 0.0837  &  8.5 \\
        UNet-32 & 8 M & 15.34\% & 15.80\% & 0.0797  &  8.5 \\
        UNet-64 & 31 M & 15.11\% & 15.56\% & 0.0795  &  8.5 \\
        UNet-96 & 70 M & 14.91\% & 15.44\% & 0.0789 &  12.5\\
        UNet-128 & 124 M &  14.29 \%& 14.75  \% & 0.0752 &  20.5 \\
        \midrule
        FNO-32-16 modes & 4 M & 8.81\% & 11.22\% & 0.0541 &  10.2 \\
        FNO-32-32 modes & 17 M & 7.47\% & 9.84\% & 0.0477 &  10.2 \\
        FNO-64-16 modes & 16 M & 6.50\% & 8.87\% & 0.0437 &  10.2\\
        FNO-64-32 modes & 67 M & 5.72\% & 7.97\% & 0.0399 &  10.2\\
        FNO-128-16 modes & 67 M & 5.49\% & 7.85\% & 0.0400 &  14.5\\
        FNO-128-32 modes & 268 M & 4.79\% & 7.19\% & 0.0373 &  14.5\\
        \midrule
        ViT-S-8 & 8 M & 9.32\% & 11.96\% & 0.0543 &  9.2\\
        ViT-S-16 & 8 M & 16.82\% & 20.85\% & 0.1047 &  8.0 \\
        ViT-B-8 & 26 M &  7.24  \% &  9.21 \% &   0.0437 &  11.8\\
        ViT-B-16 & 26 M & 14.64 \% &  18.45 \% &  0.0911 &  10.0\\
        ViT-L-8 & 87 M & 4.95 \%  & 6.86 \%   & 0.0346 &  15.2\\
        ViT-L-16 & 87 M &  11.77 \% &  15.42 \%  &  0.0731 &  12.4\\
        \midrule
        Micrometer-S-8 & 20 M & 5.32\% & 7.24\% & 0.0359  &  9.2  \\
        Micrometer-S-16 & 20 M & 5.46\% & 7.34\% & 0.0364 &  8.5\\
        Micrometer-B-8 & 70 M  & 4.39 \%  &   6.42 \%    &  0.0337   &  17.0 \\
        Micrometer-B-16 & 70 M & 5.13\% & 7.05\% & 0.0354  &  14.5\\
        Micrometer-L-8  & 292 M &  3.61 \%  & 5.95 \% & 0.0303    & 18.2\\
        Micrometer-L-16 & 292 M & 4.14\% & 6.12\% & 0.0312  &  15.5\\
        \bottomrule
    \end{tabular}
    \label{tab:different_models}
\end{table}

\begin{figure}[h!]
    \centering
    \includegraphics[width=0.8\linewidth]{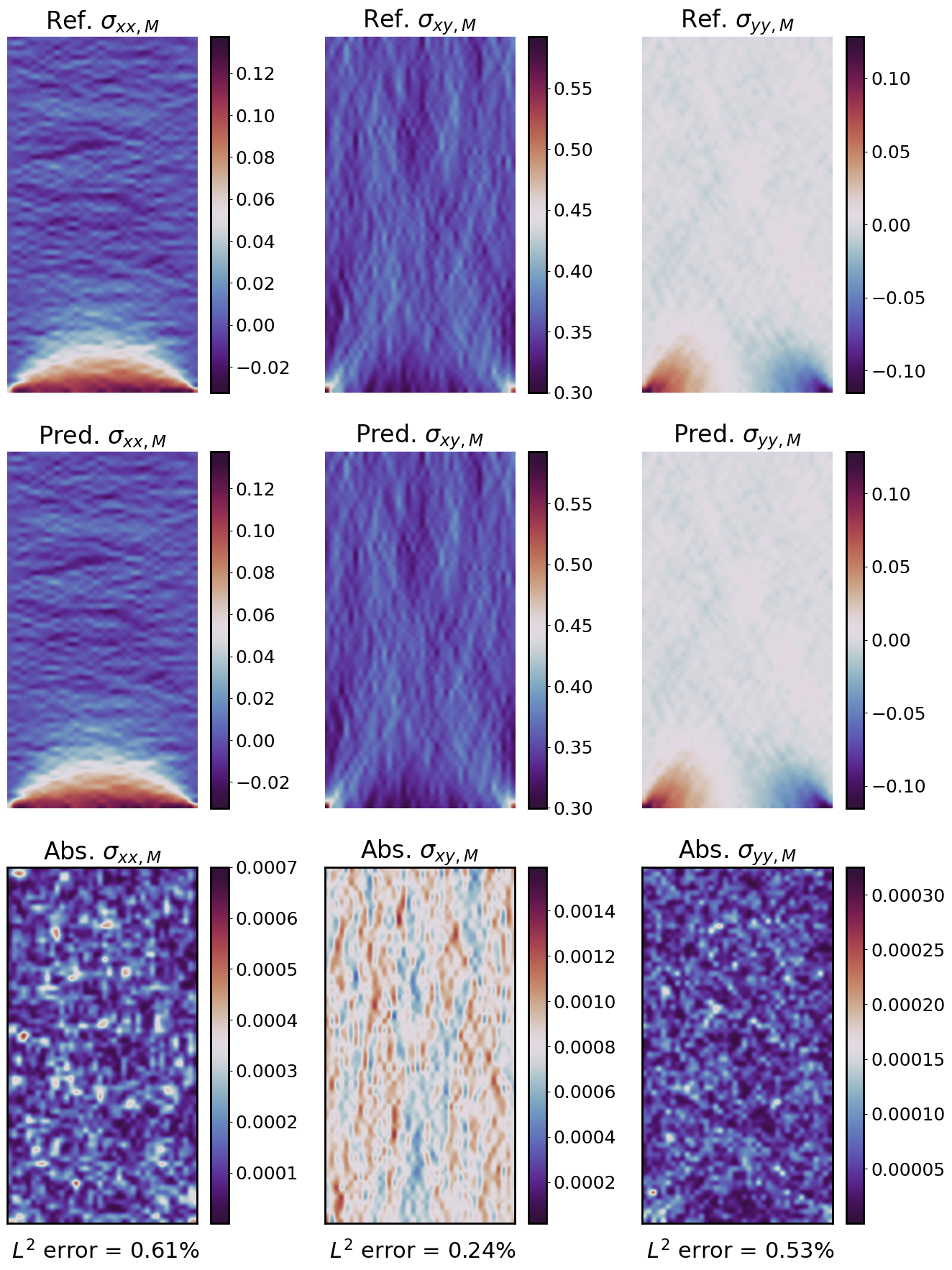}
    \caption{{\em Concurrent Mutiscale Modeling:} 
  Comparison of macrostress for a representative composite plate with low-Young's modulus ratio between FE-FFT and FE-Micrometer.}
    \label{fig:macro_pred_low}
\end{figure}

\begin{figure}[h!]
    \centering
    \includegraphics[width=0.8\linewidth]{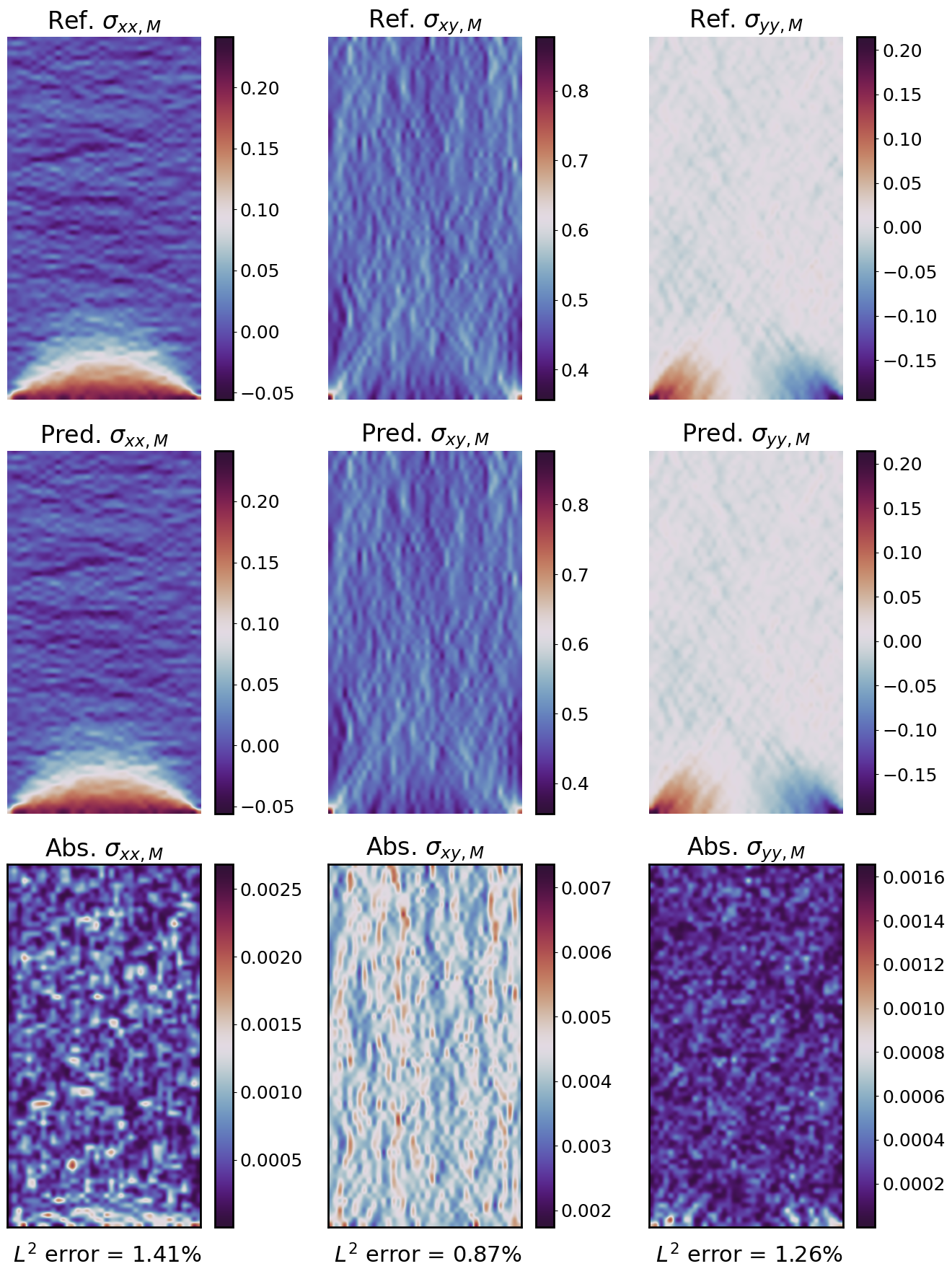}
   \caption{{\em Concurrent Mutiscale Modeling:} 
  Comparison of macrostress for a representative composite plate with medium-Young's modulus ratio between FE-FFT and FE-Micrometer.}
    \label{fig:macro_pred_medium}
\end{figure}

\end{document}